\documentclass[a4paper,11pt]{article}
\pdfoutput=1 % if your are submitting a pdflatex (i.e. if you have
\usepackage{jheppub}
\usepackage{xcolor}% for details on the use of the package, please
\usepackage[T1]{fontenc} % if needed
\usepackage{bbold}
\usepackage{amsmath}
\usepackage{amssymb}
\usepackage{mathtools}
\usepackage{stmaryrd}
\usepackage{tensor}

\usepackage{subcaption}
\usepackage{graphicx}
\usepackage{stmaryrd}
\usepackage{booktabs}

\let\<=\langle
\let\>=\rangle

\def\be#1\ee{\begin{align}#1\end{align}}

\newcommand{\al}[1]{\begin{align}#1\end{align}}
\newcommand{\spl}[1]{\begin{split}#1\end{split}}

\newcommand{\ga}[1]{\begin{gathered}#1\end{gathered}}

\usepackage{tikz}
\usetikzlibrary{calc,decorations.markings}
\tikzset{every picture/.style={line width=0.8pt}}

\tikzset{graph-1/.style = {
  line cap = round,
   line join = round,
     > = triangle 45,
     x=0.7cm, y=0.7cm,
      every node/.append style = {inner ysep=2mm}
                        }
    }% end of tikzset
    
\usetikzlibrary{matrix,decorations.pathreplacing}

\title{
Massive deformations of supersymmetric Yang-Mills matrix models%\\
}

\author[a]{Adrien Martina}
\affiliation[a]{Fields and Strings Laboratory, Institute of Physics, Ecole Polytechnique Federale de Lausanne (EPFL),
CH-1015 Lausanne, Switzerland}
\abstract{
We systematically classify all supersymmetry-preserving mass deformations of SYM matrix models 
in all dimensions ($D=3,4,6,10$). In \(D = 10\), the polarized IKKT model emerges as the only possible deformation. In \(D = 4\), we identify two massive models without a sign problem, making them attractive candidates for non-perturbative numerical studies.
\vspace{3cm}
}

\begin{document} 
\maketitle
\flushbottom

\newpage

\section{Introduction}

Among the most revolutionary ideas in modern theoretical physics, holography suggests that gravity could emerge from a fundamental gauge theory description \cite{Maldacena:1997re,tHooft:1993dmi,Aharony:1999ti,Itzhaki:1998dd,Witten:1998qj,Gubser:1998bc}. A particularly intriguing application arises in matrix quantum-mechanical models like BFSS and BMN \cite{Berenstein:2002jq,Banks:1996vh}, where gravitational phenomena are expected to emerge from a large-$N$ quantum mechanical system, avoiding the subtleties of quantum field theories.

In this work, however, we focus on an even more minimal setting where we believe gravity can emerge: super-Yang-Mills (SYM) matrix models. We define this class of matrix models by taking pure super-Yang–Mills theory in dimensions $D = 3, 4, 6 \text{ or }10$ and dimensionally reducing it all the way to a single point (zero dimension). We then consider deformations that preserve $SU(N)$ gauge invariance and the full set of original supersymmetries (possibly in a deformed form).

The most studied holographic model in this class is the IKKT model \cite{Ishibashi_1997}. Originally introduced as a non-perturbative description of type IIB string theory, it was later reinterpreted in terms of the usual decoupling limit \cite{Ooguri:1998pf}. Another interesting holographic model is the polarized IKKT model \cite{Bonelli_2002,Hartnoll:2024csr}, a mass-deformed version of the IKKT model. A key advantage is that it possesses a tunable coupling, and infinitely many well-defined observables. It has many saddles, each of which is expected to be dual to a specific ten-dimensional Euclidean gravitational background \cite{Komatsu:2024bop,Komatsu:2024ydh}. %This model will be the main focus and central interest of our study. \JP{Is this true? }
%Note that these holographic matrix models should be distinguished from the matrix models that serve as duals to low-dimensional gravity (see, e.g., \cite{Douglas:1989ve,Brezin:1990rb,Gopakumar:2022djw,Collier:2023cyw}).

Let us now discuss the structure of these matrix models. Upon dimensional reduction, the $SU(N)$ SYM gauge fields reduce to a set of $D$ Hermitian (traceless) $N \times N$ matrices $X_I$, one for each spacetime direction, and their supersymmetric partners reduce to a set of $\mathcal N = 2(D-2)$  fermionic Hermitian (traceless) $N \times N$ matrices $\psi_\alpha$. %The number of fermions and the number of bosons is related by $\mathcal N = 2 (D-2)$. 
Since the matrices no longer have any coordinate dependence, the path integral reduces to a finite number of matrix integrals. Namely, correlation functions and the partition function take the form
\begin{equation}
    \langle f(X, \psi) \rangle = \frac{1}{Z} \int [dX_I] [d \psi_\alpha] f(X,\psi) e^{-S} \,, \qquad Z=\int [d X_I] [d \psi_\alpha] e^{-S} \,, \label{eq: matrix Model observables}
\end{equation}
% The definition of these models will be given below. One of the earliest examples of such a model which is believed to be holographic is the IKKT model \cite{Ishibashi_1997}. \AM{add why IKKT < pIKKT} Recently, we have been interested in its mass-deformed variant—the polarized IKKT matrix model—which has been proposed as a dual to a specific Euclidean gravitational background \cite{Komatsu:2024bop}. This is not to be confused with matrix models dual to low dimensional gravity. [...]
% Let us now define the class of models we are considering. The SYM matrix models should thought of as pure super Yang-Mills in any dimension ($D=3,4,6,10$), dimensionally reduced down to a point, plus any deformation that is consistent with $SU(N)$ gauge invariance and supersymmetry.
% ence the path integral reduces to a finite number of matrix integrals. To be more precise, we consider a set of $D$ Hermitian $N \times N$ matrices $X_I$, one for each spacetime direction, and a set of $\mathcal N$ Hermitian $N \times N$ fermionic matrices $\psi_\alpha$. In the SYM models we are considering, $N_f = 2 (D-2)$ and $D$ can only take the values $D=3,4,6,10$.\footnote{This will be clearer where we write the action, but supersymmetry relies on specific Fierz identities that do not work out in dimensions different than those mentioned.} Correlation functions and the partition function take the form
% \begin{equation}
    % \langle f(X, \psi) \rangle = \frac{1}{Z} \int [dX_I] [d \psi_\alpha] f(X,\psi) e^{-S} \,, \qquad Z=\int [d X_I] [d \psi_\alpha] e^{-S} \,.
% \end{equation}
where $S$ is the action of the model under consideration and the measure of integration is defined in appendix \ref{app: matrix integral conventions}. For the integral to be convergent, we work in Euclidean signature.
% For our class of models, it will take the form
% \begin{equation}
    % S = \frac{1}{g_\mathrm{YM}^2} \mathrm{Tr}\left( - \frac{1}{4} [X_I, X_J]^2 + ... \right)
% \end{equation}
% namely, they are models inherited from dimensional reduction of pure SYM $D=3,4,6,10$, that we allow to deform in a manner consistent with $U(N)$ gauge invariance and supersymmetry.

These models stand out for their striking simplicity and surprising power. In the context of holographic models in ten dimensions ($D = 10$), evaluating these integrals should, in principle, capture gravitational dynamics in the large-$N$ limit.

However, this simplicity is not without cost. Even though the models look simple and well-defined, carrying out analytical or numerical computations presents some practical difficulties. A first complication is that, even though there are finitely many matrix integrals (at fixed $N$), 10 bosonic and 16 fermionic matrices still result in a large number of variables. Another complication is the so-called \textit{sign problem}. In all models we consider, the action $S$ is quadratic in the fermionic variables. When considering bosonic observables, the fermionic integral can be performed exactly, resulting in the Pfaffian of an $\mathcal N (N^2-1) \times \mathcal N (N^2-1)$ matrix $\mathcal{M}(X)$. The expectation value of a bosonic observable $f(X)$ then takes the form
\begin{equation}
    \langle f(X) \rangle = \frac{1}{Z} \int [d X_I] f(X)\mathrm{Pf} \mathcal{M}(X) e^{-S_\mathrm{bos}} \,,\qquad Z = \int [d X_I] \mathrm{Pf} \mathcal{M}(X) e^{-S_\mathrm{bos}}  \,,
    \label{eq: bosonic observables with Pfaffian}
\end{equation}
where $S_\mathrm{bos}$ is the bosonic part of $S$. Unfortunately, in $D=10$, the Pfaffian $\mathrm{Pf}  \mathcal{M} (X)$ is not positive semi-definite. This complicates the use of numerical approaches such as Monte Carlo simulations \cite{Krauth:1998xh,Fujii:2013sra}. Furthermore, this means that observables such as $\mathrm{Tr} (X_I X_I)^p$ are not necessarily positive, obstructing the use of techniques like the bootstrap \cite{Lin:2024vvg,Lin:2020mme}. It is worth noting, however, that promising results for dealing with the sign problem have been achieved through approaches such as the complex Langevin method and the Lefschetz thimble technique \cite{Fujii:2013sra,Anagnostopoulos:2020xai,Anagnostopoulos:2017gos,Anagnostopoulos:2022dak,Chou:2025moy}.
% Unfortunately, it turns out that the Pfaffian $\mathrm{Pf} \mathcal{M}(X)$ can be negative for $D=10$, hence $\mathrm{Pf} \mathcal{M}(X) e^{- S_\mathrm{bos}(X)}$ cannot be thought as a probability measure for the random variables $X_I$ since it is not positive everywhere. This leads to complications in numerical approaches such as Monte-Carlo, and for example, does not ensure that $\mathrm{Tr} (X_I X_I)^p$ is positive leading to the impossibility of applying bootstrap techniques.

For these reasons, it is valuable to explore simplified toy models that retain the essential features of the polarized IKKT model but are more tractable. 
In this work, we classify all such models. Specifically, we identify all possible mass deformations of the pure SYM matrix models $S_0$ in dimensions $D=3,4,6,10$ such that the deformed actions
\begin{equation}
    S = S_\mathrm{0} + \frac{1}{g_\mathrm{YM}^2} \mu S_1 + \frac{1}{g_\mathrm{YM}^2} \mu^2 S_2 + \dots 
\end{equation}
preserve $SU(N)$ gauge symmetry and are invariant under $\mathcal N = 2,4,8,16$ (deformed) supersymmetries, respectively. We restrict our attention to models whose actions have a real bosonic part. A similar classification was performed in the context of matrix quantum mechanics \cite{Kim:2006wg}, focusing on supersymmetric mass deformations of SYM quantum mechanics (including the BFSS model). Let us emphasize that the models we construct are not obtained by dimensional reduction of these theories. In particular, reducing the BMN model to 0+0 dimensions breaks part of the supersymmetry. A key point is that, in these mass-deformed matrix quantum mechanics, the supersymmetry transformations are explicitly time-dependent.
% The main idea underlying this classification is discussed in section \ref{app: classification idea}.

The models obtained from our classification are summarized in table \ref{the table}. We refer to these as \textit{massive SYM matrix models}.
\begin{table}[htb]
\centering
\small
{\renewcommand{\arraystretch}{1.20}%
\begin{tabular}{@{}lccccc@{}}
\toprule
{$\begin{array}{c}\text{Massive SYM}\\[-2pt]\text{matrix models}\end{array}$} &
$\mathcal{N}$ &
{$\begin{array}{c}\text{Residual}\\[-2pt]\text{symmetry}\end{array}$} &
Superalgebra &
{$\begin{array}{c}\text{Deformation}\\[-2pt]\text{parameter}\end{array}$} &
{$\begin{array}{c}\text{Pfaffian}\\[-2pt]\text{positivity}\end{array}$}\\
\midrule[0.9pt] % thicker header/body rule
$D=10$         & 16 &
  {$\begin{array}{c} SO(2,1) \times SO(7)\\[-2pt] SO(3) \times SO(6,1) \end{array}$} &
  {$\begin{array}{c} F(4)\\[-2pt] F^1(4) \end{array}$} &
  $\mu$ & No  \\
\addlinespace[0.0ex]
\midrule[0.2pt] % light full-width separator

$D=6$ type I   &  8 &
  $SO(2,1)\times SO(3)$ &
  $D(2,1,\alpha =\frac{1}{2})$ &
  $\mu$ & No  \\
\addlinespace[0.0ex]
\midrule[0.2pt]

$D=6$ type II  &  8 &
  {$\begin{array}{c} SO(5)\\[-2pt] SO(4,1) \end{array}$} &
  {$\begin{array}{c} OSp^{*}(2|4)\\[-2pt] OSp^{*}(2|2,2)\end{array}$} &
  $\mu$ & No  \\
\addlinespace[0.0ex]
\midrule[0.2pt]

$D=4$ type I   &  4 &
  $SO(3,1)$ &
  $OSp(1|2)\oplus OSp(1|2)$ &
  $\mu$ & Yes \\
\addlinespace[0.0ex]
\midrule[0.2pt]

$D=4$ type II  &  4 &
  {$\begin{array}{c} SO(3) \\[-2pt] SO(2,1) \end{array}$} &
  {$\begin{array}{c} SU(2|1) \\[-2pt] SU(1,1|1) \end{array}$} &
  $\mu_1,\mu_2$ & Yes \\
\addlinespace[0.0ex]
\midrule[0.2pt]

$D=3$          &  2 &
  $SO(2,1)$ &
  $OSp(1|2)$ &
  $\mu_1,\mu_2$ & No  \\
\bottomrule
\end{tabular}}%
\caption{Classification of massive super-Yang--Mills matrix models}
\label{the table}
\end{table}

The quantities $\mu,\mu_1, \mu_2$ are deformation parameters of mass dimension one. The number $D$ denotes the number of bosonic matrices $X_I$ while $\mathcal{N}$ denotes the number of fermionic matrices $\psi_\alpha$, and correspondingly the number of supersymmetries. We refer to $D$ as the \textit{dimension} of the matrix model, since we interpret $X^I$ as $D$-dimensional target space coordinates. Although we will soon work in Euclidean signature, the fermions satisfy Lorentzian reality conditions, as discussed in section \ref{sec: IKKT}. For this reason, the models are labeled by their Lorentzian symmetry groups. A mass deformation typically reduces the original $SO(D-1,1)$ Lorentz symmetry to a subgroup. The residual subgroup is indicated in the third column. Our notation for the superalgebras in the fourth column follows \cite{VanProeyen:1999ni}. The underlying simple super-Lie algebras were first classified by Kac \cite{Kac:1977em,Kac:1977qb}. In the last column, we indicated whether the Pfaffian is positive semi-definite in Euclidean signature.

The $D=4$ models stand out as especially promising toy models. These are the only models that do not suffer from the sign problem. Therefore, standard Monte Carlo or bootstrap methods should be applicable. It is worth noting that they still have a rich structure. For instance, the type II model, like the polarized IKKT model, possesses many saddle points.

There is an important caveat: once mass-deformed, pure SYM matrix models cannot be recovered. In every mass-deformed model we have studied, the partition function blows up as the mass parameter is taken to zero, even though the undeformed partition function is known to converge. This discontinuity arises from the fermionic mass terms, a point first noted in \cite{Austing:2001ib} and reviewed in \cite[App.B]{Komatsu:2024ydh}.

Finally, we note that a particular subset of the models constructed in this paper can be derived from SYM theories defined on the $n$-sphere $S^n$, by extending the results of \cite{Blau:2000xg} to the case $n=0$. This was first noted in the case of massive SYM quantum mechanics with $n=1$ \cite{Bobev:2024gqg}. In this identification, the mass parameter plays the role of the inverse of the sphere radius. This map between SYM on $S^n$ and the matrix models will be analyzed case by case in the main text.

This paper is structured as follows.
Section \ref{sec: IKKT} revisits pure SYM matrix models, detailing convergence properties and summarizing a few known analytic results. Section \ref{app: classification idea} presents the main idea underlying the classification of the mass-deformed models. Section \ref{sec: pIKKT} reviews the polarized IKKT model, outlining its saddle-point structure and some known analytic results.
Sections \ref{sec: 6d models}-\ref{sec: 3d model} present the new models we construct in dimensions $D = 6, 4, 3$ and provide a few analytic results in the low-mass and large-mass limits. The systematic classification of all models, the $D=4$ Pfaffian positivity proof, and other technical derivations are collected in the appendices.

\section{Review of massless SYM matrix models}
\label{sec: IKKT}
This section reviews the pure (massless) SYM matrix models in all dimensions. In particular, we examine the partition functions and observables, discuss convergence properties, and present some $SU(2)$ analytic results already established in the literature.

Pure SYM matrix models can exist only in $D=2,3,4,6,10$ because pure SYM does not exist in other dimensions. To see this, one can either argue by counting physical degrees of freedom \cite[Sec.2.4.3]{Tong:SupersymmetricFT}, or more technically by checking that the invariance of the action under supersymmetry relies on the vanishing of a term cubic in the fermions. This term vanishes only thanks to Fierz identities that are valid in these specific dimensions. However, the partition function in $D=2$ is zero because the Pfaffian \eqref{eq: bosonic observables with Pfaffian} vanishes identically.
%\JP{What about the mass deformations? Maybe massive $D=2$ is non-trivial?}\AM{addressed in a footnote in the the next section. There is no mass deformation in $D=2$ as well.}

In $D=3,4,6,10$ Euclidean dimensions, the model consists of $D$ bosonic matrices $X_I$ and $\mathcal N = 2(D-2)$ fermionic matrices $\psi_\alpha$. These are Hermitian traceless $N \times N$ matrices governed by the action\footnote{In $D=6$, where we work with two Majorana-Weyl symplectic fermions $\psi_{a=1,2}$, one should replace $\bar \psi (\ldots) \eta \to \epsilon^{a b} \bar \psi_b (\ldots) \eta_a$ everywhere in the discussion as we detail in section \ref{sec: 6d models}.}
\begin{equation}
\begin{split}
    S_\mathrm{0} = \frac{1}{g_\mathrm{YM}^2} \operatorname{Tr} & \Biggl[-\frac{1}{4} [X_I,X_J]^2 - \frac{i}{2} \bar{\psi}_\alpha \Gamma^{I}_{\alpha \beta} [X_I,\psi_\beta] \Biggr] \, . \label{eq: S IKKT}
\end{split}
\end{equation}
The gamma matrices $\Gamma^I$ satisfy the $SO(D)$ Clifford algebra. The fermion conjugate is defined by $\bar \psi \equiv \psi^\top \mathcal{C}$, where $\mathcal{C}$ is the conjugation matrix. Since the Euclidean model is obtained from Wick rotating the Lorentzian model, fermions $\psi$ obey the reality conditions of $SO(D-1,1)$. For example, in the case $D=10$, the spinors are Majorana-Weyl. Note that the Majorana condition is not invariant under Euclidean $SO(10)$ rotations, as we discuss in appendix \ref{app: gamma matrix conventions and identities}. However, this issue is resolved after integrating out the fermions, since the resulting Pfaffian is polynomial in $X_I$ and can safely be analytically continued \cite{Austing:2001ib,Krauth:1998xh}. In fact, supersymmetry would be lost if we used $SO(10)$ Majorana fermions instead of $SO(9,1)$ Majorana-Weyl. We refer the reader to appendix \ref{app: gamma matrix conventions and identities} for explicit constructions of $\mathcal{C}$, $\Gamma^I$ and $\psi$.

The model possesses $SO(D)$ rotational symmetry, as well as $SU(N)$ gauge invariance, which acts as global transformations $X_I \to U X_I U^\dagger, \psi_\alpha \to U \psi_\alpha U^\dagger$. Furthermore, it has $\mathcal  N = 2 (D-2)$ real supersymmetries. An arbitrary supersymmetry transformation $\delta_\epsilon$ can be parameterized by a spinor parameter $\epsilon$ with $\mathcal N$ degrees of freedom. Explicitly, the transformation acts on matrices as
\al{
\spl{
\delta_\epsilon X_I = \bar{\epsilon} \Gamma^I \psi \,, \qquad \qquad
\delta_\epsilon \psi = \frac{i}{2} \Gamma^{I J} \epsilon [X_I, X_J] \,. \label{eq: IKKT undeformed susy}
}
}
In our convention, $\epsilon$ commutes with everything, while $\delta$ is a Grassmann-odd operator, i.e. it anticommutes with fermions.

Bosonic observables take the form \eqref{eq: bosonic observables with Pfaffian}. The Pfaffian $\mathrm{Pf} \mathcal{M} (X)$ is taken over a matrix $\mathcal{M}(X)$ which has size $\mathcal N (N^2-1) \times \mathcal N (N^2-1)$. Explicit formulas for this matrix can be found in \cite{Krauth:1998xh}. Direct numerical evaluation of random 
$X_I$ configurations reveals that $\mathrm{Pf}\mathcal{M}(X)$ has alternating sign in the cases $D=3,6,10$. However, this is not the case for $D=4$, which led Krauth, Nicolai and Staudacher to the conjecture that the $D=4$ Pfaffian is always positive semi-definite \cite{Krauth:1998xh}. This was later proven in \cite{Ambjorn:2000bf}. We review the proof in appendix \ref{app: Pfaffian positivity}.

Since the bosonic action $S_\mathrm{bos} = - \frac{1}{4} \mathrm{Tr}[X_I,X_J]^2$ vanishes for commuting (e.g. diagonal) matrices, one may be worried that the partition function $Z$ in \eqref{eq: bosonic observables with Pfaffian} diverges. To address this, let us first consider the purely bosonic model $S_\mathrm{bos}$. Even though it stretches to infinity, the region in which $e^{-S_\mathrm{bos}}$ stays $\mathcal{O}(1)$ can still have finite volume, provided it narrows rapidly enough at large distances.\footnote{For a toy integral that mimics this feature, refer to \cite[Sec.2.1]{Komatsu:2024ydh}.} This means that the bosonic integral $\int [d X_I] e^{- S_\mathrm{bos}}$ does not necessarily diverge. After the careful treatment of \cite{Austing:2001bd,Austing:2001ib}, one obtains the following convergence properties for bosonic correlators of the $SU(N)$ bosonic model in $D\geq 3$ dimensions,
\begin{equation}
    (D=3, N\geq 4) \text{ or } (D=4, N \geq 3) \text{ or } (D\geq5,N\geq2) \iff Z_\mathrm{bos.} < \infty \,,
\end{equation}
\begin{equation}
    k < 2(D-2) N - 3D +4 \iff \langle \mathrm{Tr} (X_I  X_I)^{k/2}\rangle_{\mathrm{bos. }} < \infty  \,,
\end{equation}
where in the second line we are excluding the values of $D$ and $N$ for which $Z_\mathrm{bos.}$ diverges. In fact, any polynomial in $X$ of degree less than the critical value given above has convergent expectation value. In the large-$N$ limit, this means that all observables of finitely many matrices are well-defined.

In the supersymmetric case, one needs to additionally account for the Pfaffian. The treatment of \cite{Austing:2001pk,Austing:2001ib} then yields
\begin{equation}
    D = 4,6,10 \implies Z < \infty \,,
\end{equation}
\begin{equation}
    k< 2 (D-3) \implies \langle \mathrm{Tr} (X_I  X_I)^{k/2}\rangle < \infty \,,
\end{equation}
for any $N\geq2$.\footnote{Note that here the implication is only one way. It is not proven that all other values of $k$ or $D$ lead to divergences. In particular, the case $D=3$ is subtle. For even $N$, it turns out that the partition function trivially vanishes by examining the Pfaffian. It is conjectured to vanish for all values of $N$, not just even $N$ \cite{Krauth_1998}.} Note that contrary to the bosonic case, not all observables are necessarily convergent in the large-$N$ limit. Fermions effectively enlarge the valleys at infinity. This is analogous to the BFSS case, where the bosonic BFSS model has discrete spectrum \cite{Simon:1983jy}, but it becomes continuous once adding fermions \cite{deWit:1988xki}.\footnote{We are grateful to J. Penedones for pointing out this nice analogy.}

The closed form of the partition function of the $SU(N)$ pure SYM matrix model, first conjectured in \cite{Green:1997tn} and later computed using localization techniques \cite{Moore:1998et} takes the form
\begin{equation}
    Z = (2 \pi)^{\frac{D (N^2-1)}{2}} \mathcal{F}_N\begin{cases}
        0 & D=3 \\ \frac{1}{N^2} & D=4 \\ \frac{1}{N^2} & D=6 \\ \sum_{m| N} \frac{1}{m^2} & D=10
    \end{cases} \,,
\end{equation}
where
\begin{equation}
    \mathcal{F}_N \equiv \frac{(2 \pi)^{\frac{N-1}{2}}}{\sqrt{N}\prod_{k=1}^{N-1} (k!)} \,.
\end{equation}
The sum $\sum_{m|N}$ runs over all divisors of $N$ and the prefactor, determined in \cite{Krauth:1998xh}, has been adapted to our conventions (see appendix \ref{app: matrix integral conventions}). In the IKKT case ($D = 10$), the integral ceases to be analytic in $N$. Instead, it can be rewritten using the divisor function $\sigma_2(N) = \sum_{m\mid N} m^{2}$ \cite[Sec.2.1]{Komatsu:2024ydh}.  The properties of $\sigma_2(N)$ have been analyzed in the mathematical literature \cite{hardy1979introduction}.

Bosonic correlators in the supersymmetric theory have been extensively studied for $SU(2)$ and any $D$ \cite{Krauth:1999rc}. Specifically, all convergent moments of the form $\langle \mathrm{Tr}X_1^{2p} \rangle$ where $X_1$ is one of the $D$ bosonic matrices have been computed, yielding
\begin{equation}
    \langle \mathrm{Tr} X_1^{2} \rangle_{D=6} = \frac{1}{\sqrt{\pi}} \,, \qquad \langle \mathrm{Tr} X_1^{4} \rangle_{D=6} = \frac{25}{32} \,,
\end{equation}
\al{
\ga{
\langle \mathrm{Tr} X_1^2 \rangle_{D=10} = \frac{16}{25\sqrt{\pi}} \,, \qquad \langle \mathrm{Tr} X_1^4 \rangle_{D=10} = \frac{9}{40} \,, \qquad\langle \mathrm{Tr} X_1^6 \rangle_{D=10} = \frac{3}{8 \sqrt{\pi}} \,, \\
\langle \mathrm{Tr} X_1^8 \rangle_{D=10} = \frac{297}{1024} \,, \qquad \langle \mathrm{Tr} X_1^{10} \rangle_{D=10} = \frac{1089}{1024 \sqrt{\pi}} \,, \qquad\langle \mathrm{Tr} X_1^{12} \rangle_{D=10} = \frac{184041}{81920} \,,
}
}
while they all diverge in $D=3,4$. Note that we have adapted the results to our normalization conventions (see appendix \ref{app: matrix integral conventions}).

\section{Classification of massive SYM matrix models}
\label{app: classification idea}
In this section, we provide the main idea that allows us to classify massive SYM matrix models. Here we will assume that we have one Majorana or Majorana-Weyl spinor, meaning $D=3,4,10$. In $D=6$, we have more than one Majorana spinor so we will treat this case separately (see appendix \ref{app: proof in d=6}). Note that the $D=2$ pure SYM matrix model does not admit any mass deformation. This is because one can neither write a fermionic mass term nor a Myers term in $D=2$. The constraints that we will derive soon then imply that all deformed quantities have to vanish. %We start from the IKKT model as described in the previous section \eqref{eq: S IKKT}.
% \begin{equation}
%     S _0= \frac{1}{g_\mathrm{YM}^2} \mathrm{Tr} \left( - \frac{1}{4} [X_I, X_J]^2 - \frac{i}{2} \bar \psi \Gamma^I [X_I, \psi] \right) \, 
% \end{equation}
%It is important to note that the parameter $g_\mathrm{YM}$ has mass dimension 2, while the matrices $X_I$ have dimension $1$ and $\psi_\alpha$ have dimension $3/2$. 
% This model has maximal $\mathcal{N} = 2(D-2)$ real supersymmetries, where the supersymmetry transformation reads
% \al{
% \spl{
% \delta_\epsilon X^I & = \bar{\epsilon} \Gamma^I \psi \,, \\
% \delta_\epsilon \psi & = \frac{i}{2} \Gamma^{I J} \epsilon [X_I, X_J] \,. \label{eq:SUSY_32}
% }
% }
%The supersymmetry transformations \eqref{eq: IKKT undeformed susy} are parameterized by a parameter $\epsilon$ which has mass dimension $-1/2$.

Consider a massive deformation of $S_0$ \eqref{eq: S IKKT} parameterized by a mass parameter $\mu$,
\begin{equation}
    S = S_0 + \frac{1}{g_\mathrm{YM}^2} ( \mu S_1 + \mu^2 S_2 + \dots) \label{eq: new_S} \,.
\end{equation}
Since $g_\mathrm{YM}$ has mass dimension 2, dimensional analysis implies that $S_1,S_2,\dots$ must have mass dimension $3,2,\dots$. Noting that $X_I$ has mass dimension 1 and $\psi_\alpha$ has dimension $3/2$, the only gauge invariant combinations of $X$'s and $\psi$'s that form dimension 3 and dimension 2 operators are
\begin{equation}
    S_1 \supset \mathrm{Tr} \bar \psi M \psi, \ S_{I J K} \mathrm{Tr} X^I X^J X^K \,,
\end{equation}
\begin{equation}
    S_2 \supset   S_{I J}\mathrm{Tr} X^I X^J \,,
\end{equation}
while there are no gauge-invariant polynomial combinations that can give lower-dimensional operators. Note that $\mathrm{Tr} X^I$ vanishes for traceless matrices. On the other hand, if one allows the matrices to have a trace, the term $\mathrm{Tr} X_I$ can always be removed by completing the square with the trace contribution coming from $S_{I J} \mathrm{Tr} X_I X_J$. Inspired by  \cite{Myers:1999ps}, we refer to $S_{I JK} \mathrm{Tr} X^I X^J X^K$ as the Myers term.

The supersymmetries \eqref{eq: IKKT undeformed susy} may also need to be deformed in order to leave \eqref{eq: new_S} invariant for all $\epsilon$. We consider the generic form
\al{
\spl{
\delta_\epsilon X^I & = \bar{\epsilon} \Gamma^I \psi + \mu \bar \epsilon (\delta_B^{(1)} X^I) + \mu^2 \bar \epsilon (\delta_B^{(2)} X^I) + \ldots  \,, \\
\delta_\epsilon \psi & = \frac{i}{2} \Gamma^{I J} \epsilon [X_I, X_J] + \mu (\delta_F^{(1)} \psi) \epsilon + \mu^2 (\delta_F^{(2)} \psi) \epsilon + \ldots \,. \label{eq:SUSY_32_generic}
}
}
Note that $\epsilon$ has mass dimension $-1/2$. Thus, $\delta_B^{(1)} X^I,\delta_B^{(2)} X^I,\ldots$ must be fermionic operators of dimension $1/2,-1/2,\ldots$. Since our fermions have dimension $3/2$, we cannot build such operators and thus $\delta_B^{(k)} = 0$. On the other hand, $\delta_F^{(1)} \psi$ must be a bosonic operator of dimension 1. The only possibility is $\delta_F^{(1)} \psi \sim X_I$ while $\delta_F^{(k >1)} = 0$. Putting everything together, the most general possibility at this point is
\al{
\spl{
S = S_0 + \frac{1}{g_\mathrm{YM}^2}\mathrm{Tr} \Biggl\{\mu \bar \psi M \psi + \mu S_{I J K} X^I X^J X^K + \mu^2 S_{I J} X^{I}X^{J} \Biggr\} \,, \label{eq:Deformed S}
}
}
with the supersymmetry transformations given by
\al{
\spl{
\delta_\epsilon X^I & = \bar{\epsilon} \Gamma^I \psi \,, \\
\delta_\epsilon \psi & = \frac{i}{2} \Gamma^{I J} \epsilon [X_I, X_J] + \mu F^I \epsilon X_I \,, \label{eq:Deformed SUSY}
}
}
where $M$ and $F^I$ are matrices living in the space spanned by fully antisymmetrized products of gamma matrices, while $S_{I J K}$ is a cyclic tensor and $S_{I J}$ is a symmetric tensor. Since we are interested in actions with real bosonic part, we require that $S_{IJ} \in \mathbb R$ and $S_{I J K}^* = S_{K J I}$. 

Furthermore, we require supersymmetry, $\delta_\epsilon S = 0$ for all $\epsilon$. This is of course satisfied for the undeformed model $(\mu = 0)$, and takes the form
\begin{equation}
    \delta_\epsilon S = \mu (\delta_\epsilon S)^{(1)} + \mu^2 (\delta_\epsilon S)^{(2)} \, .
\end{equation}
We thus get two sets of constraints, $(\delta_\epsilon S)^{(1)} = 0$ and $(\delta_\epsilon S)^{(2)} = 0$. They can be written respectively as
\al{
\spl{
0 = \mathrm{Tr} \Biggl\{ [\bar \psi, X^I] (i \Gamma^I F^J \epsilon - i M \Gamma^{I J} \epsilon) X^J + 3 S_{K I J} \bar \psi X^I \Gamma^K \epsilon X^J \Biggr\} \,, \label{eq:Constr_1 Trace}
}
}
\begin{equation}
    0 = -2 \mathrm{Tr} \Biggl \{\bar \psi (M F^I - S_{I J} \Gamma^J) \epsilon X_I \Biggr \} \,. \label{eq:Constr_2 Trace}
\end{equation}
The first constraint cannot be satisfied unless $\bar \psi X^I \to \frac{1}{2} [\bar \psi, X^I]$ which requires $S_{K I J}$ to be not only cyclic, but fully antisymmetric. From \eqref{eq:Constr_1 Trace} and \eqref{eq:Constr_2 Trace} we obtain
\begin{equation}
    (i \Gamma^{[I} F^{J]} - i M \Gamma^{I J} + \frac{3}{2} S_{K I J} \Gamma^K)\epsilon = 0 \,, \label{eq:Constr_1}
\end{equation}
\begin{equation}
    (- M F^I + S_{I J} \Gamma^J)\epsilon = 0 \, .\label{eq:Constr_2}
\end{equation}
%Note that in the case of Majorana-Weyl spinors $(D=10)$, these constraints are to be imposed on the corresponding Weyl sub-space, i.e. one should multiply these two equations by $\frac{1}{2}(1+ \Gamma_*)$ on the right.\footnote{That's also the case for $D=6$ but the equations will be slightly different.} That's it. 
This needs to be imposed for all spinors $\epsilon$, which are Majorana or Majorana-Weyl depending on the dimension. In order to proceed with solving these equations, one needs to specify the dimension $D$. The detailed analysis is carried out for each $D$ in appendix \ref{app: systematic classification}. The resulting models are discussed in the next sections and summarized in table \ref{the table}.

\section{Ten-dimensional matrix model: polarized IKKT} 
\label{sec: pIKKT}
The polarized IKKT model, originally introduced in \cite{Bonelli_2002} and recently examined in \cite{Hartnoll:2024csr, Komatsu:2024bop, Komatsu:2024ydh,Hartnoll:2025ecj}, is a mass-deformed version of the IKKT model \cite{Ishibashi_1997} that retains invariance under sixteen (deformed) supersymmetries. We consider this model to be the most interesting among all the matrix models, because of its richness as a holographic description of ten-dimensional geometries \cite{Komatsu:2024bop,Komatsu:2024ydh,Hartnoll:2024csr}. In appendix \ref{app: d=10 proof}, we show that it is the unique mass deformation of IKKT that has sixteen supersymmetries and preserves $SU(N)$ gauge invariance, up to redefinition of $\mu$ and $SO(10)/(SO(3) \times SO(7))$ rotations. Note that we restricted to real bosonic deformations.

This model features ten Hermitian bosonic matrices $X_{I=1,\ldots,10}$ and sixteen Hermitian matrices of fermions $\psi_{\alpha = 1,\ldots,16}$. These are governed by the action
\begin{equation}
\begin{split}
    S_\mathrm{pIKKT} = \frac{1}{g_\mathrm{YM}^2} \operatorname{Tr} & \Biggl[-\frac{1}{4} [X_I,X_J]^2 - \frac{i}{2} \bar{\psi} \Gamma^{I} [X_I,\psi] \\ & + \frac{3 \mu^2}{2^6} X_i X_i + \frac{\mu^2}{2^6} X_p X_p + i \frac{\mu}{3} \epsilon^{i j k} X_i X_j X_k - \frac{1}{8} \mu \bar{\psi} \Gamma^{1 2 3} \psi \Biggr] \, . 
\end{split}
\end{equation}
It preserves invariance under $SU(N)$ gauge transformations, which act as global transformations $X_I \to U X_I U^\dagger, \ \psi_\alpha \to U \psi_\alpha U^\dagger$. The rotational $SO(10)$ symmetry is broken down to $SO(3) \times SO(7)$. We use the indices $i,j,\ldots=1,2,3$ to denote the directions transforming under $SO(3)$ and $p,q,r,\ldots=4,\ldots,10$ the directions transforming under $SO(7)$. The capital indices run over all values $I,J,\ldots=1,2,\ldots,10$. The matrices $\Gamma^I$ are $32 \times 32$ Hermitian matrices satisfying the $SO(10)$ Clifford algebra. One should think of this model as defined by Wick-rotating the Lorentzian model. Thus, the fermions are Majorana-Weyl. Hence, although we write them in 32-component form, the lower 16 components are identically zero. The conjugate $\bar \psi$ is defined as $\bar \psi = \psi^\top \mathcal{C}$ where $\mathcal{C}$ is the conjugation matrix of $SO(10)$. Explicit expressions for the matrices $\mathcal{C}$ and $\Gamma^I$ can be found in appendix \ref{app: gamma matrix conventions and identities}.

The mass-deformed model still possesses sixteen dynamical supersymmetries, but their form is modified relative to the original IKKT model. They act on matrices as
\al{
\spl{
\delta_\epsilon X^I & = \bar{\epsilon} \Gamma^I \psi \,, \\
\delta_\epsilon \psi & = \frac{i}{2} \Gamma^{I J} \epsilon [X_I, X_J] + \frac{3 \mu}{8} \Gamma^{1 2 3} \Gamma^i \epsilon X_i + \frac{\mu}{8} \Gamma^{1 2 3} \Gamma^p \epsilon X_p \,,
}
}
where $\epsilon$ is a Majorana-Weyl spinor (commuting) parameter, and $\delta_\epsilon$ is an anti-commuting (Grassmann-odd) operator. The supersymmetries together with the global $SO(3) \times SO(7)$ form either a gauged $F(4)$ or a gauged $F^1(4)$ superalgebra in Lorentzian signature.\footnote{The anticommutator of two supersymmetries takes exactly the form expected from $F(4)$ as was shown in \cite{Komatsu:2024ydh}.} This model can be obtained from SYM on $S^n$ in the limit $n \to 0$ \cite{Blau:2000xg}. In the notation of \cite{Blau:2000xg}, polarized IKKT arises from the family B, with $(d=10,n=0)$. Note that as in IKKT, $g_\mathrm{YM}$ can be scaled out of the integral, by rescaling $X_I = \sqrt{g_\mathrm{YM}} \tilde X_I$ and $\psi_\alpha = g_\mathrm{YM}^{3/4} \tilde \psi_\alpha$. This leaves only $\Omega \equiv \mu / \sqrt{g_\mathrm{YM}}$ as a mass parameter.

The local minima of the polarized IKKT integral are given by configurations satisfying
\begin{equation}
    \psi = 0, \qquad X_p = 0 \,, \qquad X_i = \frac{3 \mu}{8} L_i \,, \qquad [L_i, L_j] =  i \epsilon_{ijk}L_k \,.
\end{equation}
By the last condition, $L_i$ needs to form an $N$-dimensional representation of the $SU(2)$ Lie algebra, which can be reducible. The number of (inequivalent) saddles is thus given by $p(N)$, the number of partitions of $N$. For a given saddle parametrized by some irreducible representation $L_i$, the on-shell action takes the value
\begin{equation}
    S_\mathrm{pIKKT}^\mathrm{on-shell} = - \frac{9 \mu^4}{2^{13} g_\mathrm{YM}^2}\mathrm{Tr} L_i L_i \,.
\end{equation}
In particular, the dominant saddle which maximizes $\mathrm{Tr} L_i L_i$ is the $N$-dimensional irreducible representation of $SU(2)$.

The knowledge of these minima is useful for two reasons. The first reason is the limit $\Omega^2 =\mu^2/g_\mathrm{YM} \to \infty$, in which one can use a saddle point approximation to compute the observables, and they all reduce to evaluating fluctuations around the irreducible saddle \cite{Hartnoll:2024csr}. The second reason is that using supersymmetric localization methods, one can reduce the partition function and protected observables to sums over all the saddles. In particular, this aligns well with the conjecture that each saddle point should be dual to specific gravitational solutions, and indeed a one-to-one map is proposed in \cite{Komatsu:2024ydh,Komatsu:2024bop}.

Let us now discuss observables, in dimensionless units where the only tunable parameter is $\Omega \equiv \mu / \sqrt{g_\mathrm{YM}}$. Firstly, the fixed $N$, $\Omega 
%= \mu/\sqrt{g_\mathrm{YM}} 
\to 0$ limit is accessible analytically (at leading order). In this case, one can integrate out the off-diagonal components. This allows us to compute a large set of gauge invariant observables in the strict massless limit as shown in \eqref{eq: observables massless limit}.
%yielding (for the $U(N)$ theory)
%\begin{equation}
%    \langle f(\tilde X_I, \tilde \psi_\alpha) \rangle_{\mathrm{\Omega \to 0}} = \frac{\int dr_a^I d \theta_a^\alpha f(\frac{r_I}{\Omega},\frac{\theta_\alpha}{\sqrt{\Omega}}) e^{- S_0}}{\int dr_a^I d \theta_a^\alpha  e^{- S_0}} + \mathrm{subleading}
%\end{equation}
In particular, this analysis yields
\al{
\ga{
Z_{SU(N), \mathrm{pIKKT}}  \xrightarrow{\Omega \to 0} \frac{(2 \pi)^{5 (N^2-1)}}{\prod_{k=1}^N (k!)} \left( \frac{2^9}{3 \sqrt{3} \Omega^2} \right)^{N-1} \,,\\
\langle \mathrm{Tr} X_1^{2p} \rangle \xrightarrow{\Omega \to 0} N \left(\frac{2^6 (N-1)}{3 \Omega^2 N} \right)^p \frac{\Gamma(p+\frac{1}{2})}{\sqrt{\pi}} \,,\\
\langle \mathrm{Tr} X_4^{2p} \rangle \xrightarrow{\Omega \to 0} N \left(\frac{2^6 (N-1)}{\Omega^2 N} \right)^p \frac{\Gamma(p+\frac{1}{2})}{\sqrt{\pi}}\,.
}
}
Note that these observables all diverge, whereas in IKKT we know a few of them converge. This discontinuity is due to fermion mass terms, as was first mentioned in \cite{Austing:2001ib} in a similar setup and we reviewed in \cite{Komatsu:2024ydh}.

The large $\Omega$ behavior $\Omega \gg N^{1/4}$ was obtained in \cite{Hartnoll:2024csr} yielding
\begin{equation}
Z_{U(N),\mathrm{pIKKT}} \overset{\Omega \to \infty}{\longrightarrow} \frac{(2 \pi)^{5 N^2 + (N-1)/2}}{\prod_{k=1}^{N-1} k!} \frac{2^9}{3 \sqrt{3} N \sqrt{N} \Omega^2} \prod_{J=1}^{N-1} \left( \frac{3J + 2}{3 J + 1} \right)^3 e^{\frac{9 \Omega^4}{2^{15}} (N^3-N)} \,.
\end{equation}
The above computations unfortunately lie outside the region of validity of supergravity. With the use of supersymmetry, localization methods can however significantly simplify the (analytic and numerical) computation of protected observables of the form $\langle f (X_3 - i X_{10}) \rangle$ at arbitrary values of $\Omega$ and $N$, and in particular access the supergravity regime. Its result gives significant support for the conjecture that each saddle is dual to a geometry, since it is possible to match equations from both sides for each saddle \cite{Komatsu:2024ydh}. The localization result has also been recently analyzed in the statistical sense in \cite{Hartnoll:2025ecj}.

\section{Six-dimensional matrix models}
\label{sec: 6d models}
In six dimensions, we consider six Hermitian bosonic matrices $X_{I=1,\ldots,6}$, one set of four left-handed fermionic Hermitian matrices $(\psi_1)_{\alpha =1,\ldots,4}$ and another set of four left-handed fermionic Hermitian matrices $(\psi_2)_{\alpha =1,\ldots,4}$. This doubling of fermions is due to the symplectic Majorana condition that we review in appendix \ref{app: conventions in d=6}. Before discussing the mass-deformed models, we first mention the specifics of the pure SYM matrix model since $D=6$ hosts special features compared to \eqref{eq: S IKKT}. The action takes the form
\begin{equation}
    S_0^{(D=6)}= \frac{1}{g_\mathrm{YM}^2} \mathrm{Tr} \left( - \frac{1}{4} [X_I, X_J]^2 - \frac{i}{2} \epsilon^{a b} (\bar \psi_b)_\alpha \Gamma^I_{\alpha \beta} [X_I, (\psi_a)_\beta] \right) \, ,
\end{equation}
where $\epsilon^{a b}$ is the Levi-Civita tensor with $\epsilon^{1 2} = 1$. The conjugate $\bar \psi$ is defined as $\bar \psi = \psi^\top \mathcal C$ where $\mathcal{C}$ is the conjugation matrix. The matrices $\Gamma^I$ are $8 \times 8$   Gamma matrices satisfying the $SO(6)$ Clifford algebra. Recall that the fermions $\psi_i$ are left-handed. Hence, although we write them in eight-component form, the lower four components are identically zero. Also note that the action is invariant under the symplectic transformation $\psi_a \to U_{a b} \psi_b$ where $U$ is a two by two matrix satisfying $U^\top \epsilon U = \epsilon \implies \det U = 1$. Requiring that $U$ leaves the symplectic Majorana-Weyl condition \eqref{eq: symplectic condition} intact implies $U \in SU(2)$.

The supersymmetry transformations take the form
\al{
\spl{
\delta_\epsilon X^I & = \epsilon^{a b} \bar{\epsilon}_b \Gamma^I \psi_a \,, \qquad \delta_\epsilon \psi_a = \frac{i}{2} \Gamma^{I J} \epsilon_a [X_I, X_J] \,. \label{eq:SUSY in IKKT 6d}
}
}
where we have two symplectic Majorana-Weyl parameters $\epsilon_{a=1,2}$. Recall that we use the convention where $\epsilon$ is a commuting parameter while $\delta_\epsilon$ is an anticommuting operator.

In appendix \ref{app: proof in d=6} we derive all the mass deformations of $S_0^{(D=6)}$ that carry eight supersymmetries as well as $SU(N)$ gauge invariance. We find two different models that we will denote by \textit{type I} and $\textit{type II}$. The type I model breaks $SO(6) \to SO(3) \times SO(3)$, and possesses a Myers term, so that the matrix integral has many different saddles as in polarized IKKT. The type II model has $SO(5)$ global symmetry and only has a trivial saddle. 

In the following subsections, we describe these two solutions and provide a few analytical formulas.
\subsection{Type I: $ SO(6) \to SO(3) \times SO(3)$}
The action of the type I model takes the form
\begin{equation}
\begin{split}
    S_{D=6, \mathrm{type \ I}} = \frac{1}{g_\mathrm{YM}^2} \operatorname{Tr} & \Biggl[-\frac{1}{4} [X_I,X_J]^2 - \frac{i}{2} \epsilon^{a b}\bar{\psi}_b \Gamma^{I} [X_I,\psi_a] \\ & + \frac{3 \mu^2}{2^6} X_i X_i + \frac{\mu^2}{2^6} X_p X_p + i \frac{\mu}{3} \epsilon^{i j k} X_i X_j X_k - \frac{1}{8} \mu \epsilon^{a b} \bar{\psi}_b \Gamma^{1 2 3} \psi_a \Biggr] \, . \label{eq: S d=6 type I}
\end{split}
\end{equation}
The mass deformation breaks the rotational symmetry $SO(6) \to SO(3) \times SO(3)$. The supersymmetry transformations are deformed to
\al{
\spl{
\delta_\epsilon X^I & = \epsilon^{a b}\bar{\epsilon}_b \Gamma^I \psi_a \,, \\
\delta_\epsilon \psi_a & = \frac{i}{2} \Gamma^{I J} \epsilon_a [X_I, X_J] + \frac{3 \mu}{8}\Gamma^{1 2 3} \Gamma^i \epsilon_a X_i + \frac{\mu}{8} \Gamma^{1 2 3} \Gamma^p \epsilon_a X_p \,. \label{eq: susy d=6 type I}
}
}
Note that the $SU(2)$ symmetry $\psi_a \to U_{a b} \psi_b$ is preserved. In Lorentzian signature, the total global symmetry group would thus be $SO(2,1) \times SO(3) \times SU(2)$, and we have 8 supercharges. We thus expect the superalgebra to lie in the exceptional family $D(2,1,\alpha)$ \cite{VanProeyen:1999ni}. This is indeed the case as we show in appendix \ref{app: classification 6d type I}, where we find the specific value $\alpha = 1/2$ which is not isomorphic to $OSp(4|2)$. The absence of models with arbitrary $\alpha$ is surprising, as the value $\alpha=1/2$ appears arbitrary. However, it is not hard to see by allowing the coefficients in \eqref{eq: S d=6 type I} and \eqref{eq: susy d=6 type I} to take arbitrary values that supersymmetry is an algebraic constraint that leaves no free parameter except $\mu$. Thus, $\alpha = 1/2$ is, surprisingly, the unique possibility. This model can be obtained from SYM on $S^n$ with 8 supersymmetries, extrapolating \cite{Blau:2000xg} to $n = 0$. In the notation of \cite{Blau:2000xg}, it arises from the family B, with $(d=6,n=0)$.

Notice that this action has the same  numerical coefficients as polarized IKKT. Thus, the saddles (local minima of the matrix integral) have exactly the same structure. Namely, they satisfy
\begin{equation}
    \psi = 0, \qquad X_p = 0 \,, \qquad X_i = \frac{3 \mu}{8} L_i \,, \qquad [L_i, L_j] =  i \epsilon_{ijk}L_k \,.
\end{equation}
Thus, there are again $p(N)$ inequivalent saddles, each parameterized by some (in general) reducible $N$-dimensional representation of the $SU(2)$ generators $L_i$. The on-shell action takes the value
\begin{equation}
    S_{D=6, \mathrm{type \ I}}^{(\mathrm{on-shell)}} = - \frac{9 \mu^4}{2^{13} g_\mathrm{YM}^2}\mathrm{Tr} L_i L_i \,.
\end{equation}

Finally, let us discuss observables in dimensionless coordinates $(g_\mathrm{YM} \to 1, \mu \to \Omega \equiv \mu / \sqrt{g_\mathrm{YM}})$. In the strict massless limit at fixed $N$, we can integrate out off-diagonals as shown in appendix \ref{app: massless limit}. This allows us to compute a large class of observables, including
\al{
\ga{
Z_{SU(N)} \xrightarrow{\Omega \to 0} \frac{(2 \pi)^{3(N^2-1)}}{\prod_{k=1}^N (k!)} \left( \frac{2^7}{3 \sqrt{3}\Omega^2} \right)^{N-1} \,, \\
\langle \mathrm{Tr} X_1^{2 p} \rangle_{SU(N)} \xrightarrow{\Omega \to 0} N \left(\frac{2^6(N-1)}{3\Omega^{2} N}\right)^p \frac{\Gamma(p+\frac{1}{2})}{\sqrt{\pi}} \,, \\
\langle \mathrm{Tr} X_4^{2 p} \rangle_{SU(N)} \xrightarrow{\Omega \to 0} N \left(\frac{2^6(N-1)}{\Omega^{2} N}\right)^p \frac{\Gamma(p+\frac{1}{2})}{\sqrt{\pi}} \,.
}
}

\subsection{Type II: $SO(6) \to SO(5)$}
The action of the type II model takes the form
\begin{equation}
\begin{split}
    S_{D=6,\mathrm{type \ II}} = \frac{1}{g_\mathrm{YM}^2} \operatorname{Tr}  \Biggl[-\frac{1}{4} & [X_I,X_J]^2 - \frac{i}{2} \epsilon^{a b} \bar{\psi}_b \Gamma^{I} [X_I,\psi_a] \\ & + \mu^2 X_i X_i +3\mu^2 X_6^2 + 2 \mu \bar \psi_2 \Gamma^{1 2 345} \psi_1\Biggr] \, ,
\end{split}
\end{equation}
where $i=1,2,\ldots,5$ runs over the first five indices while $I,J$ run over all indices. The mass deformation breaks the rotational symmetry $SO(6)\to SO(5)$. The supersymmetry transformations are deformed to
\al{
\spl{
\delta_\epsilon X^I & = \epsilon^{a b} \bar{\epsilon}_b \Gamma^I \psi_a \,, \\
\delta_\epsilon \psi_1 & = \frac{i}{2} \Gamma^{I J} \epsilon_1 [X_I, X_J] - 3 i \mu \epsilon_1 X_6 - i \mu \Gamma^{6I} \epsilon_1 X_I \,, \\ 
\delta_\epsilon \psi_2 & = \frac{i}{2} \Gamma^{I J} \epsilon_2 [X_I, X_J] + 3 i \mu \epsilon_2 X_6 + i \mu \Gamma^{6I} \epsilon_2 X_I \,.
}
}
Note that the $SU(2)$ symmetry $\psi_a \to U_{a b} \psi_b$ with $U \in SU(2)$ remains a symmetry only if $U^\top s U = s$ where $s$ is a two by two symmetric matrix with components $s_{1 2} = s_{2 1} = 1$, $s_{1 1} = s_{22} = 0$. This restricts it to a one-parameter symmetry, infinitesimally equivalent to a $U(1)$. In Lorentzian signature, we thus have either $SO(5) \times U(1)$ or $SO(4,1) \times U(1)$ as the symmetry group, together with 8 supercharges. As we show in appendix \ref{app: classification 6d type II}, the corresponding superalgebra is either gauged $OSp^*(2|4)$ or gauged $OSp^*(2|2,2)$ respectively. This model can be obtained from SYM on $S^n$ with 8 supersymmetries, extrapolating \cite{Blau:2000xg} to $n = 0$. In the notation of \cite{Blau:2000xg}, it arises from the family A, with $(d=6,n=0)$.

Let us now discuss observables in dimensionless units $(g_\mathrm{YM} \to 1, \mu \to \Omega \equiv \mu / \sqrt{g_\mathrm{YM}})$. In the strict massless limit at fixed $N$, we can integrate out off-diagonals as shown in appendix \ref{app: massless limit}. This allows us to compute a large class of observables, including
\al{
\ga{
Z_{SU(N)} \xrightarrow{\Omega \to 0} \frac{(2 \pi)^{3(N^2-1)}}{\prod_{k=1}^N (k!)} \left( \frac{2}{\sqrt{3}\Omega^2} \right)^{N-1} \,, \\
\langle \mathrm{Tr} X_1^{2 p} \rangle_{SU(N)} \xrightarrow{\Omega \to 0} N \left(\frac{(N-1)}{\Omega^{2} N}\right)^p \frac{\Gamma(p+\frac{1}{2})}{\sqrt{\pi}} \,, \\
\langle \mathrm{Tr} X_6^{2 p} \rangle_{SU(N)} \xrightarrow{\Omega \to 0} N \left(\frac{(N-1)}{3\Omega^{2} N}\right)^p \frac{\Gamma(p+\frac{1}{2})}{\sqrt{\pi}} \,.
}
}

Since the only minimum is at $X_I = \psi_\alpha = 0$, the saddle-point analysis can be applied straightforwardly leading in particular to
\al{
\ga{
Z_{SU(N)} \xrightarrow{\Omega \to \infty} \left( \frac{16 \pi^3}{\sqrt{3} \Omega^2} \right)^{N^2-1} \,, \\
\langle \mathrm{Tr} X_1^2 \rangle_{SU(N)} \xrightarrow{\Omega \to \infty} \frac{N^2-1}{2 \Omega^2} \,, \qquad \langle \mathrm{Tr} X_6^2 \rangle_{SU(N)} \xrightarrow{\Omega \to \infty} \frac{N^2-1}{6 \Omega^2} \,.
}
}

\section{Four-dimensional matrix models}
\label{sec: 4d models}
In four dimensions, we consider four Hermitian bosonic matrices $X_{I=1,\ldots,4}$ and four fermionic Hermitian matrices $\psi_{\alpha=1,\ldots,4}$. One can either impose a Majorana or a Weyl condition on the fermions. The two are equivalent, but we work mostly in the Majorana representation. The undeformed action and supersymmetries are given in \eqref{eq: S IKKT} and \eqref{eq: IKKT undeformed susy} respectively. For more details about fermions and gamma matrices, we refer the reader to appendix \ref{app: gamma matrix conventions and identities}.

The pure SYM matrix integral in $D=4$ takes the simple form
\begin{equation}
    Z_0 = \int \left(\prod_{I=1}^4 [d X_I] \right) \mathrm{Pf} \mathcal{M}_4 (X) e^{\frac{1}{4}\mathrm{Tr} [X_I, X_J]^2} \,.
\end{equation}
where the normalization for the measure is defined in appendix \ref{app: matrix integral conventions}.
The Pfaffian can be explicitly written as
\begin{equation}
    \mathrm{Pf} \mathcal M_4 (X) = \mathrm{Pf} (-  (\mathcal{C} \Gamma^I)\otimes \mathbf X_I)  =\mathrm{det} \begin{pmatrix}
         -\mathbf X_4 + i \mathbf X_3 &  \mathbf X_2 + i \mathbf X_1\\   - \mathbf X_2 + i \mathbf X_1 & - \mathbf X_4 - i \mathbf X_3 
    \end{pmatrix}
    \,,
    \label{eq: IKKT Pfaffian formula}
\end{equation}
where $\mathbf X_I$ are $(N^2-1) \times (N^2-1)$ matrices written in the adjoint representation. Namely, the original matrices are expanded as $X_I = X_I^A T^A$ where $T^A$ are the $SU(N)$ generators in the fundamental representation normalized by $\mathrm{Tr} T^A T^B = \delta^{A B}$. The structure constants are defined by $[T_A, T_B] = i f_{ABC}T_C$. Finally, we define the $(N^2-1) \times (N^2-1)$ matrices $\mathbf X_I$, by the components $(\mathbf X_I)_{AB} \equiv f_{ABC} X_I^C$. Thus, they are real and antisymmetric. In contrast to all other dimensions, it is important to note that this determinant is always positive semi-definite. The proof was first established in \cite{Ambjorn:2000bf}. We review it in appendix \ref{app: Pfaffian positivity}. 

In appendix \ref{app: proof in d=4}, we derive all the mass deformations of the $D=4$ pure SYM matrix model that carry four supersymmetries as well as $SU(N)$ gauge invariance. We find two models that we denote by \textit{type I} and \textit{type II}. The type I model preserves $SO(4)$ symmetry, and is parameterized by a single mass parameter $\mu$. The type II model breaks $SO(4) \to SO(3)$, and it is parameterized by a continuous family of two mass parameters $\mu_1,\mu_2$. It has a Myers term, thus the matrix integral has many saddles as in polarized IKKT. The Pfaffians get deformed to functions of $\mu$ (type I) or $\mu_1$ (type II). As long as $\mu, \mu_1 \in \mathbb{R}$, the non-negativity of the Pfaffian is preserved.

\subsection{Type I: $SO(4) \to SO(4)$}
The action of the type I model takes the form
\begin{equation}
\begin{split}
    S_{D=4, \mathrm{type \ I}} = \frac{1}{g_\mathrm{YM}^2} \operatorname{Tr} & \Biggl[-\frac{1}{4} [X_I,X_J]^2 - \frac{i}{2} \bar{\psi} \Gamma^{I} [X_I,\psi] + \mu^2 X_I^2 +  \mu \bar \psi \psi \Biggr] \, .
\end{split}
\end{equation}
The mass deformation preserves the rotational $SO(4)$ symmetry. For the bosonic quadratic term to be real and positive, we restrict the mass parameter to $\mu \in \mathbb{R}$. The supersymmetry transformations are deformed to
\al{
\spl{
\delta_\epsilon X^I & = \bar{\epsilon} \Gamma^I \psi \,, \\
\delta_\epsilon \psi & = \frac{i}{2} \Gamma^{I J} \epsilon [X_I, X_J] + \mu \Gamma^I \epsilon X_I \,.
}
}
In Lorentzian signature, the symmetry group would be $SO(3,1)$, and there are 4 supercharges. In appendix \ref{app: classifiction d=4 type I}, we show that the resulting superalgebra is a (gauged) (complexified) direct sum of two simple Lie superalgebras $OSp(1|2) \oplus OSp(1|2)$. This is the supersymmetric analog to $\mathfrak{so}(3,1) \cong \mathfrak{sl}(2,\mathbb{R}) \oplus_\mathbb{C} \mathfrak{sl}(2,\mathbb{R})$. Surprisingly, this model \textit{cannot} be obtained from SYM on $S^n$ \cite{Blau:2000xg}, unlike the other matrix models.

In dimensionless units where one defines $\Omega \equiv \mu / \sqrt{g_\mathrm{YM}}$ and sets $g_\mathrm{YM} \to 1$, the $D=4$ type I integral takes the form
\begin{equation}
    Z_{D=4,\mathrm{type \ I}} = \int \left(\prod_{I=1}^4[d X_I]\right) \mathrm{Pf} \mathcal{M}_{4, \mathrm{I}}(X) e^{\mathrm{Tr}( \frac{1}{4} [X_I,X_J]^2 - \Omega^2 X_I^2)} \,,
\end{equation}
where the Pfaffian can be written as
\begin{equation}
    \mathrm{Pf} \mathcal{M}_{4,\mathrm{I}}(X) = \mathrm{Pf} \begin{pmatrix}
        0 & 2 i \Omega & - \mathbf X_3 - i \mathbf X_4 & - \mathbf X_1 + i \mathbf X_2 \\ -2 i \Omega & 0 & - \mathbf X_1 - i \mathbf X_2 & \mathbf X_3 - i \mathbf X_4 \\ -\mathbf X_3 - i\mathbf  X_4 & -\mathbf X_1 - i \mathbf X_2 & 0 & - 2i \Omega \\ -\mathbf X_1 + i \mathbf X_2 &  \mathbf X_3 - i \mathbf X_4 & 2 i \Omega & 0
    \end{pmatrix} \,.
\end{equation}
The normalization for the measure is explained in appendix \ref{app: matrix integral conventions}. The matrices $\mathbf X_I$ were defined below \eqref{eq: IKKT Pfaffian formula}. As we show in appendix \ref{app: Pfaffian positivity}, the Pfaffian is always positive semi-definite when $\Omega \in \mathbb{R}$. This implies that it can be written as the square root of a determinant. Moreover, this determinant can be simplified, yielding
\begin{equation}
    \mathrm{Pf} \mathcal{M}_{4,\mathrm{I}}(X) = \sqrt{\det(H(X)^\dagger H(X) + 4\Omega^2 \mathbb{1}_{2(N^2-1)})} \,,
\end{equation}
where
\begin{equation}
    H(X) \equiv \begin{pmatrix}
        \mathbf X_1 - i \mathbf X_2 & -\mathbf X_3 - i \mathbf X_4 \\ -\mathbf X_3 + i \mathbf X_4 & - \mathbf X_1 - i \mathbf X_2
    \end{pmatrix} \,.
\end{equation}
Note that the Pfaffian is symmetric under $\Omega \to - \Omega$ and it is an increasing function of $\Omega^2$.

Using the methods described in appendix \ref{app: massless limit}, one can compute the partition function and a few observables in the limit $\Omega \to 0$ at fixed $N$ in the $SU(N)$ theory after integrating out off-diagonal elements. This yields
\al{
\spl{
Z_{SU(N)} \xrightarrow{\Omega \to 0} \frac{(2 \pi)^{2(N^2-1)}}{\prod_{k=1}^N (k!)} \left( \frac{1}{\Omega^2} \right)^{N-1} \,, \\
\langle \mathrm{Tr} X_1^{2 p} \rangle_{SU(N)} \xrightarrow{\Omega \to 0} N \left(\frac{N-1}{N\Omega^{2}}\right)^p \frac{\Gamma(p+\frac{1}{2})}{\sqrt{\pi}} \,,
}
}
for some given matrix $X_1$ among the four matrices.

In the infinite mass limit $\Omega \to \infty$, we can perform a saddle point analysis. Since there is no Myers term, the only saddle is at $X_I = \psi_\alpha = 0$. We obtain
\al{
\spl{
Z_{SU(N)} \xrightarrow{\Omega \to \infty} \left( \frac{2 \pi}{\Omega} \right)^{2(N^2-1)} \,, \\
\langle \mathrm{Tr} X_1^2 \rangle_{SU(N)} \xrightarrow{\Omega \to \infty} \frac{N^2-1}{2 \Omega^2} \,.
}
}

%\JP{Are you being careful with the normalisation of $\mathrm{Tr} X^2$? The factors of $N$ are strange comparing the two limits...}\AM{You're right. I made a stupid mistake a long time ago, we also need to update Part II :)}

\subsection{Type II: $SO(4) \to SO(3)$}
The action of the type II model takes the form
\begin{equation}
\begin{split}
    S_{D=4, \mathrm{type \ II}} = \frac{1}{g_\mathrm{YM}^2} \operatorname{Tr} & \Biggl[-\frac{1}{4} [X_I,X_J]^2 - \frac{i}{2} \bar{\psi} \Gamma^{I} [X_I,\psi] + i \frac{2}{3} (\mu_1 + \mu_2) \epsilon_{i j k} X_i X_j X_k  \\ &   + \left( \frac{2}{9} \mu_1^2 + \frac{1}{3} \mu_1 \mu_2 \right) X_i X_i + \frac{1}{3} \mu_1 \mu_2 X_4^2 - \frac{1}{3} \mu_1 \bar \psi \Gamma^{1 2 3} \psi \Biggr] \, ,
\end{split}
\end{equation}
where $i=1,2,3$ runs over the first three indices whereas $I=1,2,3,4$ runs over all indices. The mass deformation breaks the global symmetry $SO(4) \to SO(3)$. The four supersymmetries get deformed to 
\al{
\spl{
\delta_\epsilon X^I & = \bar{\epsilon} \Gamma^I \psi \,, \\
\delta_\epsilon \psi & = \frac{i}{2} \Gamma^{I J} \epsilon [X_I, X_J] + \left( \frac{2}{3} \mu_1 + \mu_2 \right) \Gamma^{1 2 3} \Gamma^i \epsilon X_i + \mu_2 \Gamma^{1 2 3} \Gamma^4 \epsilon X_4 \,.
}
}
In addition to these symmetries, there are chiral rotations $\psi \to e^{i \alpha \Gamma_*} \psi$ that form a $U(1)$. In Lorentzian signature, the symmetry group is either $SO(3) \times U(1)$ or $SO(2,1) \times U(1)$, and there are 4 supercharges. In appendix \ref{app: classifiction d=4 type II}, we show that the corresponding superalgebra is either gauged $SU(2|1)$ or gauged $SU(1,1|1)$ respectively \cite{VanProeyen:1999ni}. Two one-parameter subfamilies of this model can be obtained from SYM on $S^n$ with 4 supersymmetries, extrapolating \cite{Blau:2000xg} to $n = 0$. In the notation of \cite{Blau:2000xg}, the one-parameter family $\mu_1 = -\mu_2$ arises from family A with $(d=4,n=0)$, while the one-parameter family $\mu_1 = 3 \mu_2$ arises from family B with $(d=4, n=0)$.

We will impose that the mass terms are positive and the Myers term is Hermitian. This restricts either $\mu_1 \mu_2 >0$, $\mu_i \in \mathbb{R}$ or $\mu_1 = - \mu_2 = i \mu$ with $\mu \in \mathbb{R}$. Note that the second possibility would yield a complex Pfaffian, so we will restrict our attention to the first case where the Pfaffian is positive.

Because of the Myers term, the $D=4$ type II matrix integral has many different saddles. The local minima satisfy
\begin{equation}
    \psi = 0 \,, \qquad X_4 = 0, \qquad X_i = \left(\frac{2}{3} \mu_1 + \mu_2 \right) L_i \,, \qquad [L_i,L_j] = i \epsilon_{i j k} \,.
\end{equation}
Thus, there are $p(N)$ inequivalent saddles, where $p(N)$ is the number of partitions of $N$. Each saddle is parameterized by some generically reducible $N$-dimensional representation $L_i$ of the $SU(2)$ generators. These minima are supersymmetric in the sense $\delta_\epsilon \psi |_\mathrm{on-shell} = 0$. The on-shell action takes the value
\begin{equation}
    S_{D=4,\mathrm{type \ II}}^{(\mathrm{on-shell})} = - \frac{\mu_2 (2 \mu_1 + 3 \mu_2)^3}{162 g_\mathrm{YM}^2} \mathrm{Tr} L_i L_i \,.
\end{equation}
In particular, the dominant saddle which maximizes $\mathrm{Tr} L_i L_i$ is the $N$-dimensional irreducible representation of $SU(2)$.

In dimensionless units $(g_\mathrm{YM} \to 1, \mu_i \to \Omega_i \equiv \mu_i / \sqrt{g_\mathrm{YM}})$, the $D=4$ type II integral takes the form
\al{
\spl{
Z_{D=4,\mathrm{type \ II}} & = \int \left(\prod_{I=1}^4[d X_I]\right) \mathrm{Pf} \mathcal{M}_{4, \mathrm{II}}(X) \ \mathrm{exp} \ \mathrm{Tr}\Biggl( \frac{1}{4} [X_I,X_J]^2 \\ & \quad - \left(\frac{2}{9} \Omega_1^2 + \frac{1}{3} \Omega_1 \Omega_2 \right)X_i^2 - \frac{1}{3} \Omega_1 \Omega_2 X_4^2 - \frac{2i}{3} (\Omega_1 + \Omega_2) \epsilon_{i j k} X_i X_j X_k\Biggr) \,.
}
}
The Pfaffian is positive semi-definite when $\Omega_1 \in \mathbb{R}$, as we prove in appendix \ref{app: Pfaffian positivity}. It can be written as a determinant,
\begin{equation}
    \mathrm{Pf} \mathcal{M}_{4,\mathrm{II}}(X) = \mathrm{det} \left(\begin{pmatrix}
         -\mathbf X_4 + i \mathbf X_3 &  \mathbf X_2 + i \mathbf X_1\\   - \mathbf X_2 + i \mathbf X_1 & - \mathbf X_4 - i \mathbf X_3
    \end{pmatrix}-\frac{2}{3} \Omega_1 \mathbb1_{2 (N^2-1)} \right) \,,
\end{equation}
where $\mathbf X_I$ are as defined below \eqref{eq: IKKT Pfaffian formula}.
%\JP{It seems that increasing $\Omega_1$ decreases the value of the det?} \AM{The overall sign is irrelevant, one could multiply what's inside the det by $(-1)$. At very large $|\Omega_1|$, we get $\mathrm{Pf} \mathcal{M}_{4,\mathrm{II}}(X) \sim \Omega_1^{2(N^2-1)}$ independently of the sign of $\Omega_1$. But around $\Omega_1 \sim 0$, the Pfaffian goes linearly in $\Omega_1$. There is no saddle at the origin, and the Pfaffian is either locally increasing or locally descreasing depending on the configuration of $X_I$. Note that it is invariant under $(X_I,\Omega) \to (-X_I, - \Omega)$.}

Integrating out off-diagonal elements as described in appendix \ref{app: massless limit}, it is possible to compute the partition function and some observables in the limit $\Omega_1, \Omega_2 \to 0$ at fixed $N$ and fixed ratio $\Omega_1 / \Omega_2$ in the $SU(N)$ theory. This yields
\al{
\ga{
Z_{SU(N)} \xrightarrow{\Omega_i \to 0} \frac{(2 \pi)^{2(N^2-1)}}{\prod_{k=1}^N (k!)} \left( \frac{3 \sqrt{3}}{\Omega_2^{1/2} (2 \Omega_1 + 3 \Omega_2)^{3/2}} \right)^{N-1} \,, \\
\langle \mathrm{Tr} X_1^{2n} \rangle_{SU(N)} \xrightarrow{\Omega_i \to 0} N \left( \frac{9}{2 \Omega_1^2 + 3 \Omega_1 \Omega_2} \frac{N-1}{N} \right)^n \frac{\Gamma(n + 1/2)}{\sqrt{\pi}} \,, \\
\langle \mathrm{Tr} X_4^{2n} \rangle_{SU(N)} \xrightarrow{\Omega_i \to 0} N \left( \frac{3}{\Omega_1 \Omega_2} \frac{N-1}{N} \right)^n \frac{\Gamma(n + 1/2)}{\sqrt{\pi}} \,.
}
}
\section{Three-dimensional matrix model}
\label{sec: 3d model}
In three dimensions, there are three Hermitian bosonic matrices $X_{I=1,2,3}$ and two fermionic Hermitian matrices $\psi_{\alpha =1,2}$. The fermions are Majorana. The undeformed action and supersymmetries are given in \eqref{eq: S IKKT} and \eqref{eq: IKKT undeformed susy} respectively. For the specifics about fermions and gamma matrices in three dimensions, we refer the reader to appendix \ref{app: gamma matrix conventions and identities}. 

In appendix \ref{app: proof in d=3}, we find a unique mass deformation of the $D=3$ pure SYM matrix model that carries two supersymmetries together with $SU(N)$ gauge invariance. This model is parameterized by two mass parameters $\mu_1, \mu_2$. The action reads
\begin{equation}
\begin{split}
    S_{D=3} = \frac{1}{g_\mathrm{YM}^2} \operatorname{Tr} & \Biggl[-\frac{1}{4} [X_i,X_j]^2 - \frac{i}{2} \bar{\psi} \Gamma^{i} [X_i,\psi] \\ & + \mu_1 \mu_2 X_i X_i + i \frac{2}{3} (\mu_1 + \mu_2) \epsilon_{i j k} X_i X_j X_k + i \mu_1 \bar \psi \psi \Biggr] \, ,
\end{split}
\end{equation}
where $i=1,2,3$. The deformation preserves the global $SO(3)$ symmetry. For the bosonic quadratic term to be real and positive, we restrict to $\mu_i \in \mathbb R, \mu_1 \mu_2 > 0$ or to $\mu_1 = - \mu_2 = i \mu$ with $\mu \in \mathbb{R}$. The two supersymmetries get deformed to
\al{
\spl{
\delta_\epsilon X^i & = \bar{\epsilon} \Gamma^i \psi \,, \\
\delta_\epsilon \psi & = \frac{i}{2} \Gamma^{i j} \epsilon [X_i, X_j] - i  \mu_2 \Gamma^i \epsilon X_i \,.
}
}
In Lorentzian signature, the bosonic symmetry group would be $SO(2,1)$, and there are 2 supercharges. In appendix \ref{app: proof in d=3}, we obtain that the superalgebra is gauged $OSp(1|2)$. The one-parameter subfamily $\mu_2 = 3 \mu_1$ can be obtained from SYM on $S^n$ with 2 supersymmetries, extrapolating \cite{Blau:2000xg} to $n = 0$. In the notation of \cite{Blau:2000xg}, it arises from the family B, with $(d=3,n=0)$.

Because of the Myers term, the $D=3$ matrix integral has many different saddles. The local minima depend on whether $\mu_1 > \mu_2$ or $\mu_1 < \mu_2$. Compactly, we can write the local minima configurations as
\begin{equation}
    \psi = 0 \,, \qquad X_i = \mathrm{max}(\mu_1, \mu_2) L_i \,, \qquad [L_i,L_j] = i \epsilon_{i j k} L_k\,.
\end{equation}
Thus, there are $p(N)$ inequivalent saddles, where $p(N)$ is the number of partitions of $N$. Each saddle is parameterized by some generically reducible $N$-dimensional representation $L_i$ of the $SU(2)$ generators. These minima are supersymmetric in the sense $\delta_\epsilon \psi |_\mathrm{on-shell} = 0$ only when $\mu_1 < \mu_2$. The on-shell action takes the value
\begin{equation}
    S_{D=3}^{(\mathrm{on-shell})} = \begin{cases}
    - \frac{1}{6} \mu_2^3 (\mu_2 - 2 \mu_1) \mathrm{Tr} L_i L_i & \mu_2 > \mu_1 \\
        - \frac{1}{6} \mu_1^3 (\mu_1 -2 \mu_2) \mathrm{Tr} L_i L_i  & \mu_1 > \mu_2
    \end{cases}
    \,.
\end{equation}
This exhibits an interesting feature. Suppose one tunes $\mu_2$ in the region $\mu_2 > \mu_1$. In the interval $2 \mu_1 > \mu_2 > \mu_1$, the dominating saddle minimizing the action is the trivial representation $L_i =0$. However, there is a transition at $\mu_2 = 2 \mu_1$ to the region $\mu_2 > 2 \mu_1$ where the dominating saddle is the irreducible representation.

Finally, let us mention observables in dimensionless units $(g_\mathrm{YM} \to 1, \mu_i \to \Omega_i \equiv \mu_i / \sqrt{g_\mathrm{YM}})$. In the strict massless limit at fixed $N$ and fixed ratio $\Omega_1 / \Omega_2$, we can integrate out off-diagonals as shown in appendix \ref{app: massless limit}. This allows us to compute a large class of observables, including e.g.
\al{
\ga{
Z_{SU(N)} \xrightarrow{\Omega_i \to 0} \frac{(2 \pi)^{\frac{3}{2}(N^2-1)}}{\prod_{k=1}^N (k!)} \left( \frac{1}{\sqrt{2}\Omega_1^{1/2} \Omega_2^{3/2}} \right)^{N-1} \,, \\
\langle \mathrm{Tr} X_1^{2 p} \rangle_{SU(N)} \xrightarrow{\Omega_i \to 0} N \left(\frac{N-1}{N\Omega_1 \Omega_2}\right)^p \frac{\Gamma(p+\frac{1}{2})}{\sqrt{\pi}} \,.
}
}

\section{Discussion}
In this work we have presented a complete classification of mass deformations of the SYM matrix models in dimensions $D = 3, 4, 6, 10$ that keep the number of supersymmetries unchanged and retain gauge invariance, as summarized in table \ref{the table}. This analysis establishes the uniqueness of the polarized IKKT model. In $D=4$ we identified two deformations with a positive semi-definite Pfaffian. The first, type I, serves as a useful benchmark, featuring a single mass parameter and only a trivial saddle, while the second, type II, has a richer structure with two mass parameters and a dominant fuzzy-sphere saddle.  Below we provide a list of interesting future directions.

\begin{itemize}
    \item The main motivation for this work was to construct numerically tractable toy models. It would be very interesting to explore whether bootstrap or Monte Carlo methods can be efficiently applied to extract new, non-perturbative information. In particular, one interesting target would be to determine whether the massive SYM matrix models become commutative at strong 't Hooft coupling ($N \to \infty$, $\lambda \equiv N / \Omega^4 \gg 1$). If this is the case, one should be able to show numerically that $\langle \mathrm{Tr} [X_I, X_J]^2 \rangle / \langle \mathrm{Tr} \{ X_I, X_J \}^2 \rangle$ becomes very small in this regime.
    \item In this paper, we derived only a partial set of analytic expressions for the new models. A full supersymmetric localization treatment along the lines of \cite{Pestun:2007rz,Asano:2012zt,Komatsu:2024ydh} remains to be developed. In particular, one could compute the partition functions, as well as protected observables of the form $\langle f(X+ i Y)\rangle$ where $X$ and $Y$ are two distinct bosonic matrices among $X_{I=1,\ldots,D}$. For example, assume $f$ is a homogeneous polynomial in $X+i Y$ of degree $k \neq 0$. This quantity would be an interesting observable when the $SO(2)$ symmetry that rotates $X$ and $Y$ is broken by the mass deformation. Otherwise, these expectation values would vanish identically as they carry a charge $k \neq 0$ under $SO(2)$. 
    % Therefore, in the $D=4$ type I model and the $D=3$ model where $SO(D)$ is preserved, only the partition function can be computed thanks to supersymmetric localization. However, in the other models which all break $SO(D)$, it would be interesting to compute e.g. the observables $\langle\mathrm{Tr}(X_1 + i X_D)^{2n}\rangle$ which would yield a non-trivial result.%\JP{Can you say a bit more? What observables do you expect to be protected in each model?}
    \item The motivation for this work was to find a variety of toy models that mimic features of the polarized IKKT model. However, the matrix models we constructed may have relevance in their own right, or at least describe some sector of models of interest. Both are true in the case of massive SYM quantum mechanics, as explained in \cite{Kim:2006wg}. These quantum-mechanical models were argued to be dual to non-critical M-theory on curved backgrounds. Moreover, some of them could be obtained from M5-brane embeddings in the 11-dimensional pp-wave geometry, or from compactifying $\mathcal N=1,2,4$ four-dimensional SYM on a three-sphere and performing a consistent truncation to the lowest modes. It would be interesting to explore whether there exist analogous setups that are dual to, or reduce to our matrix models.
    \item A subset of the matrix models could be obtained from the $n=0$ extrapolation of SYM theories on $S^n$ \cite{Blau:2000xg}. However, only one-parameter families can be extracted in this way, so the full two-parameter families of the $D = 3$ and $D = 4$ type II models cannot be derived. Moreover, the $D=4$ type I model cannot be obtained in such a way. It would be interesting to explore whether one can define SYM theories on other higher-dimensional manifolds, whose extrapolation to zero dimension reduces to the remaining matrix models.
\end{itemize}

\paragraph{Acknowledgements} We thank M. Blau, N. Bobev, S. Komatsu, J. Matos, H. Murali, J. Nishimura, J. Oliveira, J. Penedones, P. Vieira, A. Vuignier, T. Wiseman (via J. Penedones), X. Zhao for helpful discussions. This work was supported by the Simons Foundation grant 488649 (Simons Collaboration on the Nonperturbative Bootstrap) and by the Swiss National Science Foundation through the project 200020\_197160 and through the National Centre of Competence in Research SwissMAP.

\appendix
\section{Systematic classification of massive SYM matrix models}
\label{app: systematic classification}
\subsection{Uniqueness proof of polarized IKKT}
\label{app: d=10 proof}
In this section, we prove that polarized IKKT is the unique massive SYM matrix model in $D=10$ with sixteen supersymmetries and $SU(N)$ gauge invariance. 
According to the discussion in section \ref{app: classification idea}, we simply need to parameterize $M$ and $F^I$. In ten dimensions, with a Majorana-Weyl spinor $\psi$, the term $\bar \psi \Gamma^{I_1 \ldots I_r} \psi$ may be non-vanishing only for $r=3$ and its Hodge dual $r = 7$, while only even gamma matrices take left-handed spinors to left-handed spinors \cite{Polchinski:1998rr,Freedman:2012zz}. Thus, most generally\footnote{Note that we didn't consider $m_{I_1 \ldots I_7} \Gamma^{I_1 \ldots I_7}$. This is because for left-handed spinors living in the subspace where $\Gamma_* = 1$, $\Gamma^{I_1 \ldots I_7}$ and $\Gamma^{I_1 I_2I_3}$ are related.}
\begin{equation}
    M = m_{I J K} \Gamma^{I J K} \,, \qquad F^{I} = f^I + f^I_{J K} \Gamma^{J K} + f^{I}_{J K L M} \Gamma^{J K L M} \,. \label{eq: Form of F and M}
\end{equation}

We now expand the constraints \eqref{eq:Constr_1} and \eqref{eq:Constr_2} given the form of $M$ and $F^I$ \eqref{eq: Form of F and M}, and write them as a sum of antisymmetrized products of gamma matrices. To be specific let's start from \eqref{eq:Constr_1}. We write this equation in the following form,
\begin{equation}
    0 = (E_{1, KLMNP}^{I J} \Gamma^{KLMNP} + E_{2, KLN}^{I J} \Gamma^{KLN} +  E_{3,K}^{I J} \Gamma^K) \frac{1}{2} (\mathbb{1}_{32} + \Gamma_*) \, ,
\end{equation}
where we will soon specify each of the $E$'s. Since $\Gamma^{I_1 \ldots I_r}$ with $r \leq 10$ forms a complete basis of gamma matrices we obtain\footnote{Note that the term multiplying $\Gamma^{K L M N P}$ is treated specially. This is because since we are projecting this equation onto the left-handed subspace, only the ``self-dual part'' of $E_{1,KLMNP}^{I J}$ needs to vanish.},
\begin{equation}
    0 = E_{1, K L M N P}^{I J} - \frac{i}{5!} E^{I J}_{1, QRSTV} \epsilon^{QRSTV KLMNP}  \,,
\end{equation}
\begin{equation}
    0 = E_{2, K L M}^{I J} \,, \qquad \qquad 0 = E_{3,K}^{I J} \,,
\end{equation}
where $E_1$, $E_2$ and $E_3$ can be written as
\begin{equation}
    E_{1, K L M N P}^{I J} = i f^{[J}_{[K L M N} \delta_{P]}^{I]} - i m_{[KLM} \delta_{N}^I \delta_{P]}^J \,,
\end{equation}
\begin{equation}
    E_{2, K L M}^{I J} = i f_{[K L}^{[J} \delta_{N]}^{I]} + 2 i (f_{I K L N}^J - f_{J K L N}^I) + \frac{3i}{2} m_{[K L}^{\ \ \ \ [J} \delta^{I]}_{N]} \,,
\end{equation}
\begin{equation}
    E_{3,K}^{I J} = i f^{[J} \delta^{I]}_K + i (f_{I K}^{J}-f_{J K}^{I})-6i m_{K J I} + \frac{3}{2} S_{K I J} \,.
\end{equation}
The equation on $E_1$ implies
\begin{equation}
    f_{K L M N}^I = - m_{[K L M} \delta_{N]}^I  + \tilde f_{I K L M N} \,,
\end{equation}
where $\tilde f_{I K L M N}$ is a fully antisymmetric tensor which is anti self-dual, namely,
\begin{equation}
    \tilde f_{I K L M N} = \frac{i}{5!} \tilde f_{Q R S T V} \epsilon^{Q R S T V I K L M N} \, .
\end{equation}
We can then plug this in the equation for $E_2$, which implies
\begin{equation}
    f_{KL}^J = - 9 m_{K L J} \,, \qquad \tilde f_{I K L M N} = 0 \,.
\end{equation}
Finally, plugging this result in the last equation yields
\begin{equation}
    S_{I J K} = -16 i m_{I J K} \,,\qquad f^I = 0 \,.
\end{equation}
We are thus left with $S_{I J}$ and $m_{I JK}$ as unknowns.
We now consider the second set of constraints \eqref{eq:Constr_2}. The structure is analogous to before. Thus, we get
\begin{equation}
    0 = (\tilde E^I_{1,NPKLM} \Gamma^{NPKLM} + \tilde E^I_{2,K L M} \Gamma^{KLM} + \tilde E^I_{3,K} \Gamma^{K}) \frac{1}{2} (\mathbb{1} + \Gamma_*) \,.
\end{equation}
This implies
\begin{equation}
    0 = \tilde E_{1, N P K L M}^{I} - \frac{i}{5!} \tilde E^{I }_{1, QRSTV} \epsilon^{QRSTV NPKLM} \label{eq: Eq on tilde E1 in 10 d} \,,
\end{equation}
\begin{equation}
    0 = \tilde E_{2, K L M}^{I} \,, \qquad \qquad 0 = \tilde E_{3,K}^{I} \,.
\end{equation}
Plugging in $F^I$ in terms of $M$, we obtain that $\tilde{E}_1$, $\tilde{E}_2$, and $\tilde{E}_3$ read respectively
\begin{equation}
    \tilde E^I_{1,N P K L M} = 6 m_{[N P K} m_{L M]}^{\ \ \ \ \ I} + 9 m_{[ N P }^{\ \ \ \ \ Q} m^{Q}_{\ \ K L} \delta_{M]}^I \,, \label{eq: m constr 1}
\end{equation}
\begin{equation}
    \tilde E_{2, J K L}^I = 72 m_{[J K }^{\ \ \ \ \ Q} m^{Q}_{\ \ L] I} \,, \label{eq : m constr 2}
\end{equation}
\begin{equation}
     \tilde E_{3,J}^I= 36 m_{I P Q} m_{Q P J} + 6 m_{N P Q} m_{Q P N} \delta^{I J} + S^{I J} \,.
\end{equation}
The equation on $\tilde E_3$ gives the expression of $S^{I J}$ in terms of $m_{I J K}$. We are left with the two constraints from \eqref{eq: m constr 1} and \eqref{eq : m constr 2} on $m_{I J K}$. Using the constraint $\tilde E_2 = 0$, the equation on $\tilde E_1$ \eqref{eq: Eq on tilde E1 in 10 d} simplifies to
\begin{equation}
    m_{[NPK} m_{LM]I} - \frac{i}{5!} m_{[QRS} m_{TV]I} \epsilon^{QRSTV NPKLM} = 0 \,.
\end{equation}
In order to have Hermitian Myers term $S_{I J K} \in i \mathbb{R}$, it is necessary that $m_{I J K}$ is purely imaginary. Note that the fact that $m_{IJK}$ does not have \textit{both} imaginary and real parts is also necessary to have a real bosonic quadratic term. There would exist solutions where for example we set everything to 0 except some complex numbers $m_{1 2 3}, m_{23p} \in \mathbb C$ (which have both imaginary and real parts). We discard this type of solutions. Thus, both terms in the above equation have to vanish separately. This implies
\begin{equation}
    m_{[NPK} m_{LM]I} = 0 \,.
\end{equation}
Then, one can use $SO(10)$ rotations to bring $m_{I J K}$ to a form where $m_{123} \neq 0$ while among $m_{1 I J}$, one can set to 0 all terms different than $m_{1 2 3},m_{1 4 5},m_{1 6 7}$ and $m_{1 8 9}$. Thus, taking the $I = 1$, $L = 2$, $M=3$, $N=1$ component of the above equation, we obtain
\begin{equation}
    m_{1 2 3} m_{1 P K} = 0 \implies m_{1 pq} = 0 \ \ (p,q =4, \ldots,10) \,.
\end{equation}
Now that we know that $m_{1 pq} = 0$, we can consider the cases $I=1,2,3$, which imply
\begin{equation}
    m_{pqr} = 0 \,, \qquad m_{i p q} = 0 \,. \qquad (i=1,2,3 \quad p,q,r = 4,\ldots,10)
\end{equation}
All we're left with is $m_{1 2 3}$ and $m_{2 3 p}$. However, $m_{2 3 p}$ can be set to 0 thanks to rotations in the $1,p$ plane. This finishes the proof. We can thus choose (up to redefinition of $\mu$)
\begin{equation}
    m_{i j k} = - \frac{1}{8 \cdot 3!} \epsilon_{i j k} \,, \qquad (i,j,k = 1,2,3)
\end{equation}
which corresponds to
\begin{equation}
    S_{i j k} = \frac{i}{3} \epsilon_{i j k} \,, \qquad f_{J K}^I \Gamma^{J K} \epsilon X_I = \frac{3}{8} \Gamma^{1 2 3} \Gamma^i \epsilon X_i \,,\qquad f_{J K L M}^I \Gamma^{J K L M} \epsilon X_I = \frac{1}{8} \Gamma^{1 2 3} \Gamma^p \epsilon X_p \,.
\end{equation}
We thus found the only possible deformation to IKKT, namely
\begin{equation}
\begin{split}
    S = \frac{1}{g_\mathrm{YM}^2} \operatorname{Tr} & \Biggl[-\frac{1}{4} [X_I,X_J]^2 - \frac{i}{2} \bar{\psi} \Gamma^{I} [X_I,\psi] \\ & + \frac{3 \mu^2}{4^3} X_i X_i + \frac{\mu^2}{4^3} X_p X_p + i \frac{\mu}{3} \epsilon^{i j k} X_i X_j X_k - \frac{1}{8} \mu \bar{\psi} \Gamma^{1 2 3} \psi \Biggr] \, , \label{eq:action}
\end{split}
\end{equation}
supersymmetric under the transformations
\al{
\spl{
\delta_\epsilon X^I & = \bar{\epsilon} \Gamma^I \psi \,, \\
\delta_\epsilon \psi & = \frac{i}{2} \Gamma^{I J} \epsilon [X_I, X_J] + \frac{3 \mu}{8} \Gamma^{1 2 3} \Gamma^i \epsilon X_i + \frac{\mu}{8} \Gamma^{1 2 3} \Gamma^p \epsilon X_p \,. \label{eq:SUSY_32}
}
}
In \cite[App.C.1]{Komatsu:2024ydh}, we found after introducing a set of auxiliary fields that the anticommutator of two supercharges yields a sum of $SO(3)$ and $SO(7)$ generators. Thus, in Lorentzian signature, the superalgebra is either $F(4)$ (containing $SO(2,1) \times SO(7)$) or $F^1(4)$ (containing $SO(3) \times SO(6,1)$) \cite{VanProeyen:1999ni}.
\subsection{Six-dimensional classification}
\label{app: proof in d=6}
We now turn to the case of six dimensions, $D=6$. As mentioned in section \ref{app: classification idea}, the case $D=6$ is treated differently from the other dimensions since we cannot have a single Majorana spinor. Instead, we can choose our representation to be with Weyl left-handed spinors $\psi$ and $\psi^\dagger$ each of which have 4 components, for a total of 8 real degrees of freedom, or we can choose the symplectic Majorana-Weyl spinor representation where we have 2 spinors $\psi_1$, $\psi_2$ with their complex conjugates each of which are left-handed and thus have 4-components. These 4 spinors obey the reality condition
\begin{equation}
    \psi_a = \epsilon_{a b} B^{-1} (\psi_b)^* \, ,
\end{equation}
where $\epsilon^{ab}$ is the Levi-Civita tensor with $\epsilon^{1 2} = 1$. This condition effectively reduces the spinors $\{\psi_1, \psi_2, \psi_1^* , \psi_2^* \}$ to $\psi_1, \psi_2$. We will work in the symplectic Majorana-Weyl representation. Note that to go to the Weyl representation one could simply translate $\psi_2 \to \psi_1^\dagger$ and $\psi_2^\dagger \to \psi_1$ using the reality condition and call $\psi_1, \psi_1^\dagger \equiv \psi, \psi^\dagger$.
With these two spinors $\psi_1, \psi_2$, the pure SYM matrix model action reads
\begin{equation}
    S_0= \frac{1}{g_\mathrm{YM}^2} \mathrm{Tr} \left( - \frac{1}{4} [X_I, X_J]^2 - \frac{i}{2} \epsilon^{a b} \bar \psi_b \Gamma^I [X_I, \psi_a] \right) \, .
\end{equation}
Now that we have two spinors, the discussion in section \ref{app: classification idea} can be generalized and we obtain that the most general deformation in $D=6$ reads,
\al{
\spl{
S = S_0 + \frac{1}{g_\mathrm{YM}^2}  \mathrm{Tr} \Biggl\{\mu \epsilon^{a b} \bar \psi_b M \psi_a + \mu \bar \psi_a M^{a b} \psi_b + \mu S_{I J K} X^I X^J X^K + \mu^2 S_{I J} X^{I J} \Biggr\} \,,\label{eq:Deformed S 6d}
}
}
invariant under
\al{
\spl{
\delta_\epsilon X^I & = \epsilon^{a b} \bar{\epsilon}_b \Gamma^I \psi_a \,, \\
\delta_\epsilon \psi_a & = \frac{i}{2} \Gamma^{I J} \epsilon_a [X_I, X_J] + \mu F^I \epsilon_a X_I + \mu F^I_{a b} \epsilon_b X_I \,,\label{eq:Deformed SUSY 6d}
}
}
where 
\begin{equation}
    M^{a b} = M^{b a} \,, \qquad \delta^{a b} F_{a b}^I = 0 \,.
\end{equation}
Now the supersymmetry equation at order $\mu$ implies
\begin{equation}
    (i \Gamma^{[I} F^{J]} - i M \Gamma^{I J} + \frac{3}{2} S_{K I J} \Gamma^K) \frac{1}{2} (1+\Gamma_*) = 0 \label{eq:Constr_1a_6d} \,,
\end{equation}
as in other dimensions, but also,
\begin{equation}
    (i \epsilon^{c b} \Gamma^{[I} F_{c a}^{J]} - i M^{a b} \Gamma^{I J}) \frac{1}{2} (1+\Gamma_*) = 0 \label{eq:Constr_1b_6d} \,.
\end{equation}
In six dimensions the most general expansions are
\begin{equation}
    M = m_{I J K} \Gamma^{I J K} \,, \qquad M^{a b} = m^{a b}_I \Gamma^I \,,
\end{equation}
\begin{equation}
    F^I = f^I \mathbb{1} + f^I_{J K} \Gamma^{J K} \,, \qquad F^I_{a b} = f^I_{a b} + f^I_{a b; KL} \Gamma^{K L} \,,
\end{equation}
where $m_{I J K}$ is \textit{self-dual}, i.e.
\begin{equation}
    m_{I J K} = - \frac{i}{6} m_{L M N} \epsilon^{L M N I J K} \,.
\end{equation}
This is the notion of ``self-duality'' we will use in the reasoning that follows.
All along this section, when expanding the constraints, we will obtain equations of the form
\begin{equation}
    (E_{I} \Gamma^I + E_{I J K} \Gamma^{I J K}) \frac{(1 + \Gamma_*)}{2} = 0 \,,
\end{equation}
where we have included the projector $(1+\Gamma_*)/2$ to account for the fact that these equations need to be valid only on the left-handed subspace where $\Gamma_* = 1$. This turns out to be important since the rank 3 gamma matrices mix once multiplied by $\Gamma_*$, meaning $E_{I J K}\Gamma^{I J K} (1 + \Gamma_*) = (E_{I J K} - \frac{i}{3} E_{L M N} \epsilon^{L M N I J K}) \Gamma^{I J K}$. Thus, only the self-dual part of $E_{I J K}$ needs to be set to 0.
With this in mind, \eqref{eq:Constr_1a_6d} implies
\begin{equation}
    f_{[K L}^{[I} \delta^{J]}_{N]} + 6  m^{[I}_{\ \ [K L} \delta_{N]}^{J]}  - \frac{i}{3!} (f_{[PQ}^{[I} \delta^{J]}_{R]} + 6 m^{[I}_{\ \ [PQ} \delta_{R]}^{J]} ) \epsilon^{P Q R K L N} = 0 \,,
\end{equation}
\begin{equation}
    i f^{[J} \delta^{I]}_L - 2i f_{[I J] L} + m_{K M N} \epsilon^{K M N I J L} + 6 i m_{I J L} + \frac{3}{2} S_{I J L} = 0 \,.
\end{equation}
From these we obtain
\begin{equation}
    f^I = 0 \,, \qquad f^I_{J K} = - 6 m_{I J K} + f^{A}_{I J K} \,, \qquad S_{I J K} = -16 i m_{I J K} + \frac{4i}{3} f^A_{I J K}  \,,\label{eq:sol 6d 1a}
\end{equation}
where $f^A_{I J K}$ is a fully antisymmetric anti-self-dual tensor satisfying
\begin{equation}
    f^A_{I J K} = \frac{i}{6} f^A_{L M N} \epsilon^{L M N I J K} \,.
\end{equation}
In order for the Myers term to be Hermitian, $S_{I J K}$ must be purely imaginary. Note that if we don't impose Hermiticity of the Myers term, there is a solution where everything vanishes except the Myers term and $f_{I J K}^A$. The Hermiticity then constrains
\begin{equation}
    f_{I J K}^A = - 12 m_{I J K}^* \implies S_{I J K} = -16 i (m_{I J K}+ m_{I J K}^*) \,.\label{eq: 6d Myers term}
\end{equation}
The second constraint \eqref{eq:Constr_1b_6d} yields
\begin{equation}
    m_{[K}^{a b} \delta_L^{[I} \delta^{J]}_{M]} + \epsilon^{c b} f_{c a [KL}^{[I} \delta^{J]}_{M]} - \frac{i}{3!} \left(m_{[N}^{a b} \delta_P^{[I} \delta^{J]}_{Q]} + \epsilon^{c b} f_{c a [NP}^{[I} \delta^{J]}_{Q]}\right) \epsilon^{NPQKLM} = 0 \,,
\end{equation}
\begin{equation}
    2 m_{[J}^{a b} \delta^L_{I]} + \epsilon^{c b} f_{c a}^{[J} \delta^{I]}_L + 2 \epsilon^{c b} f^{[J}_{c a I] L} = 0 \,. \label{eq:sol 6d 1b}
\end{equation}
From these we obtain\footnote{Note that because of the second line, the fully antisymmetrized part over $I,J,K$ of $f^{I}_{a b J K}$ has to vanish. Thus, unlike the previous case there's no free antisymmetric anti-self-dual part.}
\begin{equation}
    f_{a b;KL}^M = - \epsilon^{a c} m_{[K}^{cb} \delta_{L]}^M \,, \qquad \qquad f_{a b}^I = - 3 \epsilon^{a c} m_I^{c b} \,.
\end{equation}
Let us now discuss the second set of constraints, coming from supersymmetry at order $\mu^2$. In this case we obtain
\begin{equation}
    (-\epsilon^{a b} M F^I - \epsilon^{c b} M F_{c a}^I - M^{a b} F^I - M^{c b} F_{c a}^I + \epsilon^{ab} S_{I J} \Gamma^J) \frac{1}{2} (\mathbb{1} + \Gamma_*) = 0 \,.
\end{equation}
Now the equations do not manifestly decouple as they did in \eqref{eq:Constr_1a_6d} and \eqref{eq:Constr_1b_6d}. After expanding in gamma matrices we obtain
\al{
\spl{
\Bigl( & - \epsilon^{c b} m_{J K N} f_{c a}^I + 6 \epsilon^{c b} m_{L [J K} f_{c a N] L}^I - m_{[J}^{a b} f_{K N]}^I - m_{[J}^{c b} f_{c a K N]}^I 
\\ & + 6 \epsilon^{a b} m_{L[J K} f_{N] L}^I \Bigr) \Gamma^{J K N}(\mathbb{1} + \Gamma_*) = 0 \,, \label{eq: 6d constr 2 Gamma3}
}
}
\al{
\spl{
& - i e^{a b} m_{J K L} f_{M N}^I \epsilon^{J K L M N P} - 6 \epsilon^{a b} m_{P K L} f_{L K}^I - i \epsilon^{c b} m_{J K L} f_{c a M N}^I \epsilon^{J K L M N P} \\ & \ - 6 \epsilon^{c b} m_{P K L} f_{c a L K}^I  - 2 m^{a b}_J f_{J P}^I - m_P^{c b} f_{c a}^I - 2 m_J^{c b} f_{c a J P}^I + \epsilon^{a b} S_{I P} = 0 \,.\label{eq: 6d constr 2 Gamma1}
}
}
Once again, only the self-dual part in $J K N$ of the term in the parentheses of the first equation needs to vanish. Using the solutions \eqref{eq:sol 6d 1a} and \eqref{eq:sol 6d 1b}, it turns out that the symmetric parts and antisymmetric parts in $a, b$ of \eqref{eq: 6d constr 2 Gamma3} decouple. That is also the case of \eqref{eq: 6d constr 2 Gamma1} once antisymmetrized over $I,P$. From the antisymmetric part in $a,b$ of \eqref{eq: 6d constr 2 Gamma3} and using \eqref{eq:sol 6d 1a} \eqref{eq: 6d Myers term} we obtain
\begin{equation}
    m_{L [J K} f_{N] L I} = 0 \implies -6m_{L [J K} m_{N] L I}  -12 m_{L [J K} m^*_{N] L I} = 0 \,. \label{eq: final eq for mIJK in d=6}
\end{equation}
Let us now distinguish cases
\subsubsection{Type I : $SO(6) \to SO(3) \times SO(3)$}
\label{app: classification 6d type I}
The first case is $m_{I J K} \neq 0$. Note that by the self-duality condition, if $m_{I J K}$ is non-vanishing, some components necessarily have non-vanishing real parts. Without loss of generality, we choose this real part to lie in $m_{1 2 3}$. One can then perform $SO(6)$ rotations of $\mathrm{Re} \ m_{I J K}$ such that one sets to 0 all components in $\mathrm{Re} \ m_{1 IJ}$ different than $\mathrm{Re \ }m_{1 2 3}$ and $\mathrm{Re \ }m_{1 4 5}$. Self-duality of $m_{I J K}$ implies that the imaginary part of $m_{I J K}$ is expressed in terms of the real part of $m_{I J K}$. One can then show by taking suitable components of \eqref{eq: final eq for mIJK in d=6}, that all components of $m_{I J K}$ other than $\mathrm{Re} \ m_{1 2 3}$ and $\mathrm{Im} \ m_{4 5 6} = - \mathrm{Re} m_{1 2 3}$ vanish. Namely we obtain (up to $SO(6)$ rotations),
\begin{equation}
    m_{1 2 3} = m \in \mathbb{R} \,,\qquad m_{4 5 6} = - i m \,,
\end{equation}
while all other components vanish.

Then, one can look at the other equations. In particular, the symmetric part in $a,b$ \eqref{eq: 6d constr 2 Gamma3} and the symmetric in $a,b$, antisymmetric in $I,P$ part of \eqref{eq: 6d constr 2 Gamma1}. More precisely, defining
\begin{equation}
    E^{a b I}_{J K N} \equiv 3m_{J K N} m_I^{a b} - 3 m_{I[ J K} m_{N]}^{a b} + 3 m_{L [J K} \delta_{N]}^I m_L^{a b} - m_{[J}^{a b} f_{K N]}^I + \epsilon^{c d} m_{[J}^{c b} m_{K}^{d a} \delta_{N]}^I \,.
\end{equation}
The equation \eqref{eq: 6d constr 2 Gamma3} implies
\begin{equation}
    E_{J K N}^{a b I} - \frac{i}{3!} E_{P Q R}^{a b I} \epsilon^{P Q R J K N} = 0 \,,
\end{equation} 
while \eqref{eq: 6d constr 2 Gamma1} implies
\begin{equation}
    F_{J K}^{a b} \equiv \frac{i}{3}m_{N P L} m_{M}^{a b} \epsilon^{N P L M K J} + 2 m_{J K L} m_L^{a b} - \frac{2}{3} m_N^{a b} f_{J K}^N + \frac{4}{3} \epsilon^{c d} m_{[J}^{c b} m_{K]}^{d a} = 0 \,.
\end{equation}
Now taking a linear combination $E^{a b N}_{J K N} - \frac{i}{3!} E^{a b N}_{P Q R} \epsilon^{PQRJKN} - F_{J K}^{a b} $ of the two equations, one obtains that the only possibility is $m^{a b}_I = 0$. For convenience we choose $m = - 1/96$. Putting everything together, we found the first solution
\begin{equation}
\begin{split}
    S = \frac{1}{g_\mathrm{YM}^2} \operatorname{Tr} & \Biggl[-\frac{1}{4} [X_I,X_J]^2 - \frac{i}{2} \epsilon^{a b}\bar{\psi}_b \Gamma^{I} [X_I,\psi_a] - \frac{1}{8} \mu \epsilon^{a b} \bar \psi_b \Gamma^{1 2 3} \psi_a \\ &  + i \frac{\mu}{3} \epsilon_{i j k} X_i X_j X_k + \frac{3 \mu^2}{2^6} X_i X_i + \frac{\mu^2}{2^6} X_p X_p\Biggr] \, ,
\end{split}
\end{equation}
which is invariant under
\al{
\spl{
\delta_\epsilon X^I & = \epsilon^{a b} \bar{\epsilon}_b \Gamma^I \psi_a \,, \\
\delta_\epsilon \psi_a & = \frac{i}{2} \Gamma^{I J} \epsilon_a [X_I, X_J] + \frac{3 \mu}{8}\Gamma^{1 2 3} \Gamma^i \epsilon_a X_i + \frac{\mu}{8} \Gamma^{1 2 3} \Gamma^p \epsilon_a X_p \,.
}
}
\paragraph{Identifying the superalgebra:}Note that this action is invariant under $SO(2,1) \times SO(3)$ rotations, and under $SU(2)$ transformations acting on the symplectic index, $\psi_a \to U_{a b} \psi_b, U\in SU(2)$. The bosonic symmetry group is thus isomorphic to $SO(4) \times Sl(2,\mathbb{R})$ which forms the bosonic part of the exceptional continuous family of superalgebras $D(2,1,\alpha)$. To ensure that $D(2,1,\alpha)$ is indeed the superalgebra and compute $\alpha$, it is important to compute the anticommutators of two supercharges. To obtain a superalgebra which closes off-shell, we will introduce a set of auxiliary fields $K_{m=1,\ldots,3}$, and consider the action
\begin{equation}
    S \to \tilde S = S + \frac{1}{2} \mathrm{Tr} K_m K_m \,.
\end{equation}
For $\tilde S$ to be supersymmetric, we add a few terms to the supersymmetry transformations,
\al{
\spl{
\delta_\epsilon X^I & = \epsilon^{ab} \bar{\epsilon}_b \Gamma^I \psi_a \,, \\
\delta_\epsilon \psi_a & = \frac{i}{2} \Gamma^{I J} \epsilon_a [X_I, X_J] + \frac{3 \mu}{8}\Gamma^{1 2 3} \Gamma^i \epsilon_a X_i + \frac{\mu}{8} \Gamma^{1 2 3} \Gamma^p \epsilon_a X_p + i \nu_a^m K_m \,, \\
\delta_\epsilon K_m & = - \epsilon^{a b} \bar \nu_b^m \Gamma^I [X_I, \psi_a] + i \frac{\mu}{4} \epsilon^{a b} \bar \nu_b^m \Gamma^{1 2 3} \psi_a \,,
}
}
where $\nu_a^m(\epsilon)$ is a spinorial parameter that will be constrained in terms of $\epsilon$ to obtain closure off-shell. This construction is inspired by \cite{Berkovits:1993hx,Asano:2012zt}. Now we can compute the action of the anticommutator $\{\delta_{\epsilon}, \delta_\eta \}$ on the relevant matrices exactly as in \cite[App.A.1]{Komatsu:2024ydh}. Imposing the closure of the algebra with the introduction of the auxiliary fields imposes conditions on $\nu_m^a$ that restrict it in terms of $\epsilon$. To obtain these constraints, it is sufficient to look at the undeformed model ($\mu = 0$). We obtain in particular
\al{
\ga{
\epsilon^{a b} \bar \eta_b \Gamma^I \nu_a(\epsilon) + (\epsilon \leftrightarrow\eta) = 0 \,,\\
\bar \epsilon_b \Gamma^J \eta_a \Gamma^J_{\alpha \beta} - (\eta_a)_\alpha (\bar \epsilon_b)_\beta - (\nu_a^m(\eta))_\alpha (\bar \nu_b^m(\epsilon))_\beta + (\epsilon \leftrightarrow \eta) = 0 \,.
}
}
Now that the constraints are known, we can compute the anticommutators for the deformed supercharges. Defining $\delta_{\epsilon} \equiv \epsilon_{\alpha a} Q_{\alpha a}$, we obtain
\al{
\spl{
\{ Q_{\alpha a}, Q_{\beta b} \} & = 2i \epsilon^{a b} (\mathcal C \Gamma^I)_{\alpha \beta} [X_I, \cdot] -i \frac{3 \mu}{8} \epsilon_{a b} (\mathcal C\Gamma^{123} \Gamma^{i j})_{\alpha \beta} M_{i j} \\ & \ \ + i \frac{\mu}{8} \epsilon_{a b}  (\mathcal C \Gamma^{1 2 3} \Gamma^{p q})_{\alpha \beta} M_{pq} + \frac{i \mu}{2} (\mathcal C \Gamma^{1 2 3})_{\alpha \beta} T_{a b} \,,
}
}
where the antisymmetric generators $M_{i j}$ and $M_{pq}$ form the $SO(3) \times SO(3)$ sub-algebra of the $SO(6)$ generators that act as
\begin{equation}
    M_{KL} X_I = - i (\delta_{I K} X_L - \delta_{I L} X_K) \,, \qquad M_{KL} \psi_a = - \frac{i}{2} \Gamma^{KL} \psi_a \,, \label{eq: rotations generators on matrices}
\end{equation}
whereas $T_{ab}$ are the generators of the $SU(2)$ that acts only on the symplectic index $a=1,2$ as
\begin{equation}
    T_{c d} \psi_a = \frac{i}{2} (\delta_{a c} \epsilon^{d e} + \delta_{a d} \epsilon^{c e}) \psi_e \,.
\end{equation}
Since the $SO(3)$ and $SU(2)$ algebras are isomorphic, it is useful to recast the antisymmetric generators $M_{i j}$ and $M_{pq}$ to symmetric generators taking the same form as $T_{ab}$. We first map the $SO(3)$ generators $M_{i j}$ acting on $i=1,2,3$ to $T^{(1)}_{a b} \equiv \frac{i}{2} \epsilon_{ac}\sigma_{c b}^k \epsilon_{k i j} M_{i j}$ where $\sigma^k$ are the Pauli matrices. We perform the same map for $M_{pq}$ by simply shifting the indices, namely $T^{(2)}_{a b} \equiv \frac{i}{2} \epsilon_{ac}\sigma_{c b}^r \epsilon_{r pq} M_{p q}$ where $(\sigma^4,\sigma^5,\sigma^6) \equiv (\sigma^1,\sigma^2,\sigma^3)$ and $\epsilon^{pqr}$ is the Levi-Civita tensor for the components $4,5,6$ with $\epsilon^{4 5 6} = 1$. The index $a,b=1,2$. The generators are all symmetric and all satisfy $T_{11}^\dagger = T_{2 2}, T_{12}^\dagger = - T_{1 2}$. Also, because we wrote the gamma matrices in a tensor product notation \eqref{eq: 6d weyl matrices}, it is useful to separate the spinor index $\alpha=1,\ldots,4$ into double index notation $\alpha = \alpha' \alpha''$, where $\alpha', \alpha'' = 1,2$. For example,  we defined
\begin{equation}
    \gamma^i_{\alpha \beta} \equiv(\gamma^i)_{\alpha' \alpha'', \beta' \beta''} \equiv (\sigma^3 \otimes \sigma^i)_{\alpha' \alpha'', \beta' \beta''} = \sigma^3_{\alpha' \beta'} \sigma^i_{\beta' \beta''} \,.
\end{equation}
With these notations at hand, our $SU(2)\times SU(2) \times SU(2)$ generators $T^{(1)}_{\alpha'' \beta''}, T^{(2)}_{\alpha' \beta'}, T_{ab}$ together with the supercharges $Q_{\alpha' \alpha'' a}$ form the algebra (removing the gauge transformation)
\al{
\ga{
\{ Q_{\alpha' \alpha'' a}, Q_{\beta' \beta'' b} \} = - i \frac{3 \mu}{4} \epsilon_{a b} \epsilon_{\alpha' \beta'} T^{(1)}_{\alpha'' \beta''} + i \frac{\mu}{4} \epsilon_{a b} \epsilon_{\alpha'' \beta''} T^{(2)}_{\alpha' \beta'} + i \frac{\mu}{2} \epsilon_{\alpha' \beta'} \epsilon_{\alpha'' \beta''} T_{a b} \,, \\
[T_{a b}, Q_{\alpha' \alpha'' c}] = - \frac{i}{2} (\epsilon_{a c} Q_{\alpha' \alpha'' b}+\epsilon_{b c} Q_{\alpha' \alpha'' a} ) \,, \\
[T^{(1)}_{\alpha'' \beta''}, Q_{\alpha' \gamma'' a}] = - \frac{i}{2} (\epsilon_{\alpha'' \gamma''} Q_{\alpha' \beta'' a} + \epsilon_{\beta'' \gamma''} Q_{\alpha' \alpha'' a}) \,, \\
[T^{(2)}_{\alpha' \beta'}, Q_{\gamma' \alpha'' a}] = - \frac{i}{2} (\epsilon_{\alpha' \gamma'} Q_{\beta' \alpha'' a} + \epsilon_{\beta' \gamma'} Q_{\alpha' \alpha'' a}) \,, \\
[T_{a b}, T_{c d}] = - i (\epsilon_{a b} T_{b d} + \epsilon_{b d} T_{a c}) \,,
}
}
where the last line also applies to $T^{(1)}$ and $T^{(2)}$. As a consistency check, note that the three coefficients in the anticommutators sum to zero, $-\frac{3}{4} + \frac{1}{4} + \frac{1}{2} = 0$ as required by Jacobi identities \cite{Frappat:1996pb}. Up to rescaling the supercharge, this is exactly the $D(1,2,\alpha=\frac{1}{2})$ superalgebra that can be found in the literature \cite{Ivanov:2002pc, Frappat:1996pb}.

\subsubsection{Type II : $SO(6) \to SO(5)$}
\label{app: classification 6d type II}
The second case is $m_{I J K} = 0$. In this case, it is straightforward to obtain from \eqref{eq: 6d constr 2 Gamma3} and \eqref{eq: 6d constr 2 Gamma1} that
\begin{equation}
    m_I \epsilon m_J - m_J \epsilon m_{I} = 0 \,. \label{eq: 6d type II main eq}
\end{equation}
We use a matrix product notation where $m_I$ are two by two matrices with components $m_I^{a b}$ and $(\epsilon)_{a b} = \epsilon^{a b}$. Note that all the terms in the action that don't involve $m^{a b}_I$ are invariant under $\psi_a \to R_{a b} \psi_b$ where $R$ is a symplectic 2 by 2 matrix, namely
\begin{equation}
    R^\top \epsilon R = \epsilon \implies \det R = 1 \,.
\end{equation}
One thus has two types of possible reparameterization on $m_I^{a b}$. We can rotate $I$ with fixed $a,b$ thanks to $SO(6)$ rotations, or we can transform $m_I$ with fixed $I$ by $m_I \to R^\top m_I R$. Without loss of generality, assume $m_1^{a b} \neq 0$. Using $R$, we can transform $m_1$ to make it such that $m_1^{1 1} = m_1^{22} = 0$, while $m_1^{1 2} \neq 0$. Then, using $SO(6)$, we rotate all $m^{1 2}_I$'s such that $m^{1 2}_{I=2,\ldots,6} = 0$. Finally, using \eqref{eq: 6d type II main eq}, we obtain $m^{a b}_{I=2,\ldots,6} = 0$. Thus the only non-vanishing $m_I^{a b}$ is
\begin{equation}
    m^{1 2}_1 \equiv m \neq 0 \,.
\end{equation}
The final solution is thus\footnote{We chose $m = i$.}
\begin{equation}
\begin{split}
    S = \frac{1}{g_\mathrm{YM}^2} \operatorname{Tr} & \Biggl[-\frac{1}{4} [X_I,X_J]^2 - \frac{i}{2} \epsilon^{i j} \bar{\psi}_j \Gamma^{I} [X_I,\psi_i] + 2 \mu \bar \psi_2 \Gamma^{1 2 345} \psi_1 + \mu^2 X_i X_i +3\mu^2 X_6^2\Biggr] \, ,
\end{split}
\end{equation}
invariant under
\al{
\spl{
\delta_\epsilon X^I & = \epsilon^{i j} \bar{\epsilon}_j \Gamma^I \psi_i \,, \\
\delta_\epsilon \psi_1 & = \frac{i}{2} \Gamma^{I J} \epsilon_1 [X_I, X_J] - 3 i \mu \epsilon_1 X_6 - i \mu \Gamma^{6I} \epsilon_1 X_I \,, \\ 
\delta_\epsilon \psi_2 & = \frac{i}{2} \Gamma^{I J} \epsilon_2 [X_I, X_J] + 3 i \mu \epsilon_2 X_6 + i \mu \Gamma^{6I} \epsilon_2 X_I \,.
}
}
\paragraph{Identifying the superalgebra:} This model has $SO(5)$ global symmetry, and the original $SU(2)$ symplectic symmetry of the undeformed model is broken down to a $U(1)$. The bosonic symmetry algebra is thus $SO(5) \times U(1)$. To find the superalgebra, we first close it off-shell by introducing auxiliary fields, as in the type I case (see section \ref{app: classification 6d type I}). Defining $\delta_\epsilon \equiv \epsilon_{a \alpha} Q_{a \alpha}$, we obtain that the only non-vanishing anticommutator is
\begin{equation}
    \{ Q_{1 \alpha}, Q_{2 \beta} \} = - \mu( \mathcal{C} \Gamma^{6 i j})_{\alpha \beta} M_{i j} - 2 \mu (\mathcal{C} \Gamma^{1 23 45})_{\alpha \beta} T \,,
\end{equation}
where $M_{i j}$ are $SO(5)$ rotation generators (see \eqref{eq: rotations generators on matrices}) and $T$ is a $U(1)$ generator acting only on the symplectic index of the fermions as
\begin{equation}
    T \psi_1 = \psi_1 \,, \qquad T \psi_2 = - \psi_2 \,.
\end{equation}
Moreover, the generators $T$ and $M_{i j}$ commute with each other and their commutator with $Q_{a \alpha}$ is non-vanishing. We thus identify the $OSp(2|4)$ superalgebra \cite{Freedman:1983na}. Since the rotational group in Lorentzian signature becomes either $SO(5)$ or $SO(4,1) $, we write the superalgebra as $OSp^*(2|4)$ or $OSp^*(2|2,2)$, respectively, following \cite{VanProeyen:1999ni}.
\subsection{Four-dimensional classification}
\label{app: proof in d=4}
In $D=4$, we have Majorana \textit{or} Weyl spinors. We will impose a Majorana condition. Also note that in 4d, $\{ \mathbb{1},\Gamma^I, \Gamma^{I J}, \Gamma^I \Gamma_* , \Gamma_* \}$ forms a complete basis of $4 \times 4$ matrices. However, $\bar \psi \Gamma^{I_1 \ldots I_r} \psi$ vanishes for $r=1,2$. We can thus parameterize
\begin{equation}
    M = m \mathbb{1} + m_* \Gamma_* + m_I \Gamma^I \Gamma_* \, , 
\end{equation}
\begin{equation}
    F^I = f^I_0 \mathbb{1} + f_*^I \Gamma_* +f^I_J \Gamma^J + f_{* J}^I \Gamma^J \Gamma_* +f^I_{J K} \Gamma^{J K} \,,
\end{equation}
where by construction $f_{J K}^I = - f_{K J}^I$. The first set of constraints \eqref{eq:Constr_1} gives the equations
\begin{equation}
    f^I_J = f^J_I \,,\quad (\mathbb{1}) \qquad \qquad f_{* J}^I = f_{* I}^J \,, \quad (\Gamma_*) \label{eq: d=4 id and gstar}
\end{equation}
\begin{equation}
    \frac{i}{2} f_{* K}^{[J} \epsilon^{I] K L M} - \frac{i}{2} m_* \epsilon_{I J L M} + i f_{[M}^{[J} \delta^{I]}_{L]} - i m_0 \delta_L^{[I} \delta_M^{J]} = 0 \,, \quad (\Gamma^{L M}) \label{eq: d=4 gg}
\end{equation}
\begin{equation}
    f_*^{[J} \delta_M^{I]} + f_{K L}^{[J} \epsilon^{I] K L M} + 2 m^{[J} \delta^{I]}_M = 0 \,, \quad (\Gamma^M \Gamma_*) \label{eq: d=4 ggstar}
\end{equation}
\begin{equation}
    i f_0^{[J} \delta^{I]}_L + 2i f_{[K L]}^{[J} \delta_K^{I]} - i m_K \epsilon^{K I J L} + \frac{3}{2}S_{L I J} = 0 \,, \quad (\Gamma^L) \label{eq: d=4 g}
\end{equation}
The first two equations \eqref{eq: d=4 id and gstar} \eqref{eq: d=4 gg} imply
\begin{equation}
    f^I_J = m_0 \delta^I_J \,, \quad f_{* J}^I = m_* \delta^I_J \,.
\end{equation}
Through various contractions it can be shown that equations \eqref{eq: d=4 ggstar} and \eqref{eq: d=4 g} imply, in particular, that $f^I_{J K}$ is totally anti-symmetric. Thus, it will be useful to parameterize
\begin{equation}
    S_{I J K} \equiv \epsilon_{I J K L} \tilde S_L \,, \qquad f^I_{J K} \equiv \epsilon_{I J K L} \tilde f_L \,.
\end{equation}
With these definitions, the equations \eqref{eq: d=4 ggstar} and \eqref{eq: d=4 g} imply
\begin{equation}
    f_0^I = 0 \,, \qquad\tilde f_I = \frac{m_I}{2} - \frac{3 i}{4} \tilde S_I \,,
\end{equation}
\begin{equation}
    f_*^I= -3 m_I + \frac{3i}{2} \tilde S_I \,.
\end{equation}
Finally, plugging these in \eqref{eq:Constr_2} we obtain
\begin{equation}
    m_0 \left(4 m_I - \frac{3i}{2} \tilde S_I\right) = 0 \,, \qquad (\Gamma_*) \qquad \qquad m_* \left( 4m_I - \frac{3i}{2} \tilde S_I \right) = 0 \,,\qquad (\mathbb{1})
\end{equation}
\begin{equation}
    - m_0 \epsilon^{I J K L} \tilde f_L - \frac{1}{2} m_0 m_L \epsilon^{I J K L} - 2 m_* \delta_{[J}^I\tilde f_{K]} + m_*  m_{[J} \delta^I_{K]}  = 0 \,,\qquad (\Gamma^{J K})
\end{equation}
\begin{equation}
    \epsilon^{I J K L} m_K \tilde f_L = 0 \,, \qquad (\Gamma^J \Gamma_*)
\end{equation}
\begin{equation}
    S_{I J} = f_*^{I} m_{J} +2 m_I \tilde f_J + (m_0^2 - m_*^2-2 \tilde f^L m_L) \delta_{I J} \,.\qquad (\Gamma^J)
\end{equation}
We now have two cases. If at least one of the two variables $m_*$ and $m_0$ is non-vanishing, then everything vanishes except $m_0$, $m_*$, $f^I_J$, $f_{*J}^I$ and $S_{I J}$. We thus have a two-parameter family of solutions $(m, m_*)$ and the deformation preserves $SO(4)$ symmetry. On the other hand, if both $m_0$ and $m_*$ vanish, then all these equations impose is that $m^I$ and $\tilde S^I$ point in the same direction. By an $SO(4)$ transformation, we choose $\tilde S^I = (0,0,0,i s)$ and $m^I = (0,0,0,m)$. We thus have a two-parameter deformation that breaks $SO(4) \to SO(3)$. Below are written the explicit forms of the solutions.
\subsubsection{Type I: $SO(4) \to SO(4)$} 
\label{app: classifiction d=4 type I}
The first solution preserves $SO(4)$ symmetry and its action takes the form\footnote{Compared to the above we have defined $\mu m_0 \equiv \mu_1$ and $\mu m_* \equiv i \mu_2$.}
\begin{equation}
\begin{split}
    S = \frac{1}{g_\mathrm{YM}^2} \operatorname{Tr} \Biggl[-\frac{1}{4} [X_I,X_J]^2 - \frac{i}{2} \bar{\psi} \Gamma^{I} [X_I,\psi] +  \mu_1 \bar \psi \psi + i \mu_2 \bar \psi \Gamma_* \psi + (\mu_1^2 + \mu_2^2) X_I^2 \Biggr] \, ,
\end{split}
\end{equation}
with the supersymmetry transformations given by
\al{
\spl{
\delta_\epsilon X^I & = \bar{\epsilon} \Gamma^I \psi \,, \\
\delta_\epsilon \psi & = \frac{i}{2} \Gamma^{I J} \epsilon [X_I, X_J] + \mu_1 \Gamma^I \epsilon X_I + i \mu_2 \Gamma^I \Gamma_* \epsilon X_I \,.
}
}
By field redefinition $\psi \to e^{i \alpha \Gamma_*} \psi$, with
\begin{equation}
    \alpha = \frac{1}{\sqrt{2}} \sqrt{\frac{\mu_1}{\sqrt{\mu_1^2 + \mu_2^2}} + 1} \,,
\end{equation}
one can set $\mu_2 \to 0$, $\mu_1 \to \mu \equiv \sqrt{\mu_1^2 + \mu_2^2}$. One of the parameters is thus fictitious. We obtain
\begin{equation}
\begin{split}
    S = \frac{1}{g_\mathrm{YM}^2} \operatorname{Tr} & \Biggl[-\frac{1}{4} [X_I,X_J]^2 - \frac{i}{2} \bar{\psi} \Gamma^{I} [X_I,\psi] +  \mu \bar \psi \psi + \mu^2 X_I^2 \Biggr] \, ,
\end{split}
\end{equation}
with the supersymmetries given by
\al{
\spl{
\delta_\epsilon X^I & = \bar{\epsilon} \Gamma^I \psi \,, \\
\delta_\epsilon \psi & = \frac{i}{2} \Gamma^{I J} \epsilon [X_I, X_J] + \mu \Gamma^I \epsilon X_I \,.
}
}
\paragraph{Identifying the superalgebra:} This model is invariant under $SO(4)$ rotations that act as
\begin{equation}
    \delta_\omega X^I = -2 \ \omega^{I}_{\ J} X^J \,, \qquad \qquad \delta_\omega \psi = - \frac{1}{2} \omega_{I J} \Gamma^{I J} \psi \,,
\end{equation}
where $\omega$ is an antisymmetric parameter. To identify the superalgebra, it is convenient to switch to Weyl spinors and write everything in Lorentzian signature. Namely, we write $\psi = U \psi_L + U^* \psi_L^*$ with $\psi_L = (\psi_L^1, \psi_L^2, 0 ,0)$ and $U$ a unitary matrix such that $U^\dagger \Gamma_* U = (\mathbb{1}_2, -\mathbb{1}_2)$ (see appendix \ref{app: D=4 gamma matrices}). Defining $\delta_\epsilon = (\epsilon_L)_\alpha Q_\alpha + (\epsilon_L^*)_\alpha Q_\alpha^\dagger$ and $\delta_\omega \equiv - i \ \omega^{\mu \nu} M_{\mu \nu}$, we obtain (after introducing auxiliary fields and discarding gauge transformations)
\al{
\ga{
\{ Q_\alpha, Q^\dagger_\beta \} = 0 \,, \\
\{ Q_\alpha, Q_\beta \} = -i \mu (\gamma^{\mu \nu} \epsilon)^*_{\alpha \beta} M_{\mu \nu} \,, \qquad \qquad \{ Q^\dagger_\alpha, Q^\dagger_\beta \} = i \mu (\gamma^{\mu \nu} \epsilon)_{\alpha \beta} M_{\mu \nu} \,,\\
[M_{\mu \nu}, Q_\alpha] = - \frac{i}{2} (\gamma_{\mu \nu}^* Q)_\alpha \,, \qquad \qquad [M_{\mu \nu}, Q^\dagger_\alpha] = - \frac{i}{2} (\gamma_{\mu \nu} Q^\dagger)_\alpha \,, \\
[M_{\mu \nu} , M_{\rho \sigma}] = i (\eta_{\mu \sigma} M_{\nu \rho} - \eta_{\nu \rho} M_{\sigma \nu} - \eta_{\nu \sigma} M_{\mu \rho} + \eta_{\mu \rho} M_{\sigma \nu}) \,.
}
}
We identify two separate superalgebras since $Q$ and $Q^\dagger$ anti-commute. To understand how the algebra separates, we can define $J_i = - \frac{1}{2} \epsilon_{i j k} M_{j k}$, $K_i = M_{0 i}$, $J^+_i = J_i + i K_i$, and $E_0 = J_3^+$, $E_+ = J_1^+ + i J_2^+$, $E_- = - J_1^+ + i J_2^+$. The generators $E_0$, $E_+$ and $E_-$ form a $sl(2,\mathbb{R})$ algebra given by
\begin{equation}
    [E_0, E_\pm] = \pm E_\pm \,,\qquad [E_+, E_-] = -2E_0 \,.
\end{equation}
It turns out that these are precisely what appear in the product of supercharges, namely
\begin{equation}
    \{ Q^\dagger_\alpha, Q^\dagger_{\beta} \} = 2 \mu \begin{pmatrix}
        E_- & E_0 \\ E_0 & E_+
    \end{pmatrix}_{\alpha \beta} \,,
\end{equation}
while in $\{Q_\alpha, Q_\beta\}$, one finds a similar structure with $J_i^-\equiv J_i - i K_i$ instead of $J_i^+$. The above forms the $OSp(1|2)$ algebra \cite{Frappat:1996pb}. Altogether, we thus identify the superalgebra as $OSp(1|2) \oplus OSp(1|2)$.

\subsubsection{Type II: $SO(4) \to SO(3)$} 
\label{app: classifiction d=4 type II}
The second solution breaks $SO(4) \to SO(3)$. Its action reads\footnote{Compared to the previous variables, we have defined $\mu_1 \equiv - 3 \mu m$ and $\mu_2 \equiv \mu \left( \frac{3 s}{2} + 3 m \right)$.}
\begin{equation}
\begin{split}
    S = \frac{1}{g_\mathrm{YM}^2} \operatorname{Tr} & \Biggl[-\frac{1}{4} [X_I,X_J]^2 - \frac{i}{2} \bar{\psi} \Gamma^{I} [X_I,\psi] - \frac{1}{3} \mu_1 \bar \psi \Gamma^{1 2 3} \psi \\ &  + i \frac{2}{3} (\mu_1 + \mu_2) \epsilon_{i j k} X_i X_j X_k + \left( \frac{2}{9} \mu_1^2 + \frac{1}{3} \mu_1 \mu_2 \right) X_i X_i + \frac{1}{3} \mu_1 \mu_2 X_4^2 \Biggr] \, ,
\end{split}
\end{equation}
and it is invariant under the supersymmetry transformations
\al{
\spl{
\delta_\epsilon X^I & = \bar{\epsilon} \Gamma^I \psi \,, \\
\delta_\epsilon \psi & = \frac{i}{2} \Gamma^{I J} \epsilon [X_I, X_J] + \left( \frac{2}{3} \mu_1 + \mu_2 \right) \Gamma^{1 2 3} \Gamma^i \epsilon X_i + \mu_2 \Gamma^{1 2 3} \Gamma^4 \epsilon X_4 \,,
}
}
where $i = 1,2,3 \neq 4$. In order for the Myers term to be Hermitian we need $\mu_1 + \mu_2 \in \mathbb{R}$ and for the mass terms to be positive we need $\frac{2}{3} \mu_1^2 + \frac{1}{3} \mu_1 \mu_2 >0$ and $\mu_1 \mu_2 >0$. This leaves two possibilities. The first is $\mu_1 > 0$ and $\mu_2 > 0$ and every term is manifestly fine. The only other possibility is $\mu_1 = i \mu$ and $\mu_2 = - i \mu$ which leads to a vanishing Myers term and to the action 
\begin{equation}
\begin{split}
    S_{\mu_1 = - \mu_2 = i \mu} = \frac{1}{g_\mathrm{YM}^2} \operatorname{Tr} & \Biggl[-\frac{1}{4} [X_I,X_J]^2 - \frac{i}{2} \bar{\psi} \Gamma^{I} [X_I,\psi] - \frac{i}{3} \mu \bar \psi \Gamma^{1 2 3} \psi + \frac{1}{9} \mu^2 X_i X_i + \frac{1}{3} \mu^2 X_4^2 \Biggr] \, .
\end{split}
\end{equation}
This action does not have a real Pfaffian (we would need $\mu \in i \mathbb R$ which would give an unbounded potential). We thus discard this possibility and restrict to $\mu_1, \mu_2 \in \mathbb R$ with $\mu_1 \mu_2 \geq 0$.
\paragraph{Identifying the superalgebra:} The model is invariant under $SO(3)$ rotations and $U(1)$ chiral transformations $\psi \to e^{i \alpha \Gamma_*} \psi$. After the introduction of auxiliary fields, one obtains in the Majorana representation (discarding gauge transformations)
\begin{equation}
    \{Q_\alpha, Q_\beta \} = \frac{1}{2} \left( \frac{2}{3} \mu_1 + \mu_2\right) (\mathcal{C} \Gamma^{123} \Gamma^{i j})_{\alpha \beta} M_{i j} + \left( \frac{2}{3} \mu_1 + \mu_2\right) \delta_{\alpha \beta} T \,,
\end{equation}
where $M_{i j}$ are the $SO(3)$ generators as in \eqref{eq: rotations generators on matrices} and $T$ is the infinitesimal $U(1)$ chiral transformation that only acts on fermions namely
\begin{equation}
    T X_I = 0\,, \qquad T \psi \equiv \Gamma_*\psi \,.
\end{equation}
We thus identify the $SU(2|1)$ superalgebra \cite{Ivanov:2015hfa}. In Lorentzian signature, the $SO(3)$ algebra is Wick rotated to $SO(3)$ or $SO(2,1)$. Correspondingly, the superalgebra is either $SU(2|1)$ or $SU(1,1|1)$ \cite{VanProeyen:1999ni}.

\subsection{Three-dimensional classification}
\label{app: proof in d=3}
In three dimensions, we impose the Majorana condition which implies that $\psi$ has two real components. Note that in $D=3$, $\{\mathbb{1},\Gamma^i\}$ form a complete basis of gamma matrices. We now do exactly the same ansatz as in \eqref{eq:Deformed S}, \eqref{eq:Deformed SUSY}. In this case, the fermionic matrices can be written most generically as
\begin{equation}
    M = m \mathbb{1} \,, \qquad F^I = f^I + f^I_J \Gamma^J \,.
\end{equation}
Also, since $S_{i j k}$ needs to be fully antisymmetric and we only have 3 dimensions, the only possibility is
\begin{equation}
    S_{I J K} = s \ \epsilon_{I J K} \,.
\end{equation}
The first set of constraints \eqref{eq:Constr_1} implies
\begin{equation}
    f_I^J = f_J^I \, , \qquad i \delta_K^{[I} f^{J]} + f_L^{[J} \epsilon^{I] L K} - m \epsilon^{I J K} + \frac{3}{2}s \epsilon_{K I J} = 0 \, ,
\end{equation}
from which we obtain
\begin{equation}
    f^I = 0 \,, \qquad f^I_J = \left(m -\frac{3}{2} s \right) \delta_{I J} \, .
\end{equation}
The second set of constraint \eqref{eq:Constr_2} then yields
\begin{equation}
    S_{I J} = m f^I_J = m \left(m - \frac{3 s}{2} \right) \delta_{I J} \, .
\end{equation}
The parameters $s$ and $m$ are arbitrary. Thus, we have a two-parameter family of solutions $(\mu_1, \mu_2)$ (after convenient reparametrization of $s$ and $m$). Altogether, the family of solutions is 
\begin{equation}
\begin{split}
    S = \frac{1}{g_\mathrm{YM}^2} \operatorname{Tr} & \Biggl[-\frac{1}{4} [X_i,X_j]^2 - \frac{i}{2} \bar{\psi} \Gamma^{i} [X_i,\psi] \\ & + i \mu_1 \bar \psi \psi + \mu_1 \mu_2 X_i X_i + i \frac{2}{3} (\mu_1 + \mu_2) \epsilon_{i j k} X_i X_j X_k  \Biggr] \, ,
\end{split}
\end{equation}
supersymmetric under the transformations
\al{
\spl{
\delta_\epsilon X^I & = \bar{\epsilon} \Gamma^I \psi \,, \\
\delta_\epsilon \psi & = \frac{i}{2} \Gamma^{i j} \epsilon [X_i, X_j] - i  \mu_2 \Gamma^i \epsilon X_i \,.
}
}
In particular, for $\mu_2 = 0$, the deformation fully preserves the original undeformed supersymmetries.
\paragraph{Identifying the superalgebra:} To identify the superalgebra, we simply compute anticommutators of supersymmetry transformations since in three dimensions, the off-shell closure of the superalgebra is automatic. We obtain
\begin{equation}
    \{ Q_\alpha, Q_\beta \} = -2i (\mathcal{C } \Gamma^i)_{\alpha \beta} [X_i, \cdot] + \mu_2 (\mathcal{C}\Gamma^{i j})_{\alpha \beta} M_{i j} \,,
\end{equation}
where $M_{i j}$ acts as \eqref{eq: rotations generators on matrices} on matrices and $[X_i, \cdot]$ is a field-dependent gauge transformation. The supercharges transform as spinors under $M_{i j}$. In Lorentzian signature, the $SO(3)$ rotations become $SO(2,1)$. We thus identify the superalgebra $OSp(1|2)$ \cite{hardy1979introduction}.
% To look for the saddles, let us assume $X^i = \alpha J^i$ ($\psi = 0$). Then, the action is
% \begin{equation}
%     S(\alpha) = \frac{\alpha^4}{2} - \frac{2}{3} (\mu_1 + \mu_2) \alpha^3 + \mu_1 \mu_2 \alpha^2 
% \end{equation}
% The saddles are at $\alpha = \mu_1, \mu_2$ namely
% \begin{equation}
%     X_i = \mu_1 J_i \quad \text{or} \quad X_i = \mu_2 J_i
% \end{equation}
% The second saddle, $X_i = \mu_2 J_i$ is supersymmetric (in the sense that $\delta_\epsilon \psi = 0$). The on-shell action is
% \begin{equation}
%     S =\frac{1}{3} \mu_1^3 \left(\mu_2 - \frac{\mu_1}{2}\right) \mathrm{Tr} J_i^2 \quad (X_i = \mu_1 J_i) \qquad \qquad S = \frac{1}{3} \mu_2^3 \left(\mu_1 - \frac{\mu_2}{2}\right) \mathrm{Tr} J_i^2 \qquad (X_i = \mu_2 J_i)
% \end{equation}
% The $\alpha = \mu_2$ saddle dominates (minimizes $S$) for $\mu_2 \geq \mu_1$ (otherwise the $\mu_1$ saddle dominates). For $\mu_2 \geq 2 \mu_1$, it leads to an exponentially growing $Z$.

\section{Pfaffian positivity in four dimensions}
\label{app: Pfaffian positivity}
In this appendix, we review the proof that the Pfaffian in the $D=4$ SYM matrix model is positive semi-definite \cite{Ambjorn:2000bf}, and extend it to the mass-deformed models. Switching from the Majorana to the Weyl representation as discussed in appendix \ref{app: D=4 gamma matrices}, the undeformed integral over fermions reduces to
\begin{equation}
    \int \left(\prod_{\alpha=1}^2 \prod_{A=1}^{N^2-1} d \chi_{A,\alpha}^* d \chi_{A,\alpha} \right)\mathrm{exp}(-\chi_A^\dagger (i \underbrace{f_{ABC}X_I^C}_{\equiv (\mathbf X_I)_{AB}} \bar \gamma^I) \chi_B) = \mathrm{det}(\underbrace{i \bar \gamma^I \otimes \mathbf X_I}_{\equiv A(X)}) \,,
\end{equation}
where in our representation of $\bar \gamma^I$ \eqref{eq: gbar in 4d}, the matrix $A$ takes the form
\begin{equation}
    A(X) \equiv \begin{pmatrix}
         -\mathbf X_4 + i \mathbf X_3 &  \mathbf X_2 + i \mathbf X_1\\   - \mathbf X_2 + i \mathbf X_1 & - \mathbf X_4 - i \mathbf X_3
    \end{pmatrix} \,. \label{eq: Pfaffian A(X) def}
\end{equation}
The key idea to show that $\det A$ is positive semi-definite is to use the fact that when we deal with \textit{Euclidean} gamma matrices in 4 dimensions, there exists a 2 by 2 matrix $\tilde B$ such that $\tilde B \bar \gamma^I \tilde B = - (\bar\gamma^I)^*$ and $\tilde B^2 = -\mathbb{1}_2$. In our representation this matrix is simply $\tilde B = i \sigma^2$. This implies that there exists a matrix $J = \tilde B \otimes \mathbb{1}_{N^2-1}$ such that
\begin{equation}
    J \equiv \begin{pmatrix}
        0 & \mathbb{1}_{N^2-1} \\ - \mathbb{1}_{N^2-1} & 0
    \end{pmatrix} \,, \quad J^2 = - \mathbb 1_{N^2-1}   \quad \implies J A(X) J = - A(X)^* \,. \label{eq: Pfaffian conjugation property}
\end{equation}
This property implies that if $v$ is an eigenvector of the matrix $A$ with eigenvalue $\lambda$, then $(J v)^*$ is an eigenvector with eigenvalue $\lambda^*$. Note also that $ v \neq 0 \implies(J v)^* \neq v$, so the two eigenvectors are always different. This means that every eigenvalue \textit{always} comes with another one which is its complex conjugate. Thus, the determinant is a product of many pairs $\lambda \lambda^*$ and is always positive semi-definite. One may be tempted to argue similarly for $D=6$ since the Pfaffian also reduces to a determinant in this case \cite{Krauth:1998xh}. However, in $D=6$, such a matrix $\tilde B$ does not exist.
Let us now generalize this proof to the mass-deformed models.
\paragraph{Type I:} For the type I theory the fermionic integral in the Majorana representation yields (in dimensionless coordinates $\mu \to \Omega$)
\al{
\spl{
\int \left(\prod_{\alpha=1}^4 \prod_{A=1}^{N^2-1} d \psi_{A,\alpha}\right) & \mathrm{exp}\left(- \frac{1}{2} \psi_{A \alpha} (- f_{ABC} X_C^I (\mathcal C \Gamma^I)_{\alpha \beta} + 2 \Omega \mathcal C_{\alpha \beta} \delta_{AB}) \psi_{B \beta}\right) \\ & = \mathrm{Pf} \underbrace{(-  (\mathcal{C} \Gamma^I)\otimes \mathbf X_I + 2 \Omega \mathcal C \otimes \mathbb{1}_{N^2-1})}_{\equiv \mathcal{M}_{4,\mathrm{I}}(X)} \,.
}
}
It turns out that it is again better to go to Weyl representation, but the action contains terms of the form $\chi^* i \sigma^2 \chi^*$ and $-\chi i \sigma^2\chi$. Thus the Pfaffian does not reduce to a determinant. Doing so we obtain
\begin{equation}
    \mathrm{Pf} \mathcal{M}_{4,\mathrm{I}}(X) = \mathrm{Pf} \begin{pmatrix}
        0 & 2 i \Omega & - \mathbf X_3 - i \mathbf X_4 & - \mathbf X_1 + i \mathbf X_2 \\ -2 i \Omega & 0 & - \mathbf X_1 - i \mathbf X_2 & \mathbf X_3 - i \mathbf X_4 \\ -\mathbf X_3 - i\mathbf  X_4 & -\mathbf X_1 - i \mathbf X_2 & 0 & - 2i \Omega \\ -\mathbf X_1 + i \mathbf X_2 &  \mathbf X_3 - i \mathbf X_4 & 2 i \Omega & 0
    \end{pmatrix} \,.
\end{equation}
%Note that since $\mathrm{Pf}(\ldots)^2 = \mathrm{det(\ldots)}$, we can always write $\mathrm{Pf}(\ldots) =\pm \sqrt{\det (\ldots)}$ locally. Also, it is not hard to see that the Pfaffian is real when $\Omega$ is real. The question is whether $\det(\ldots)$ ever crosses zero. If so, the sign could change. However, b
To argue that the Pfaffian is positive, we will use the property $(\mathrm{Pf} \mathcal{M})^2 = \mathrm{det} \mathcal M$.
By simple manipulations (rearranging columns and changing some signs) one obtains that
\al{
\spl{
\mathrm{det} & \begin{pmatrix}
        0 & 2 i \Omega & - \mathbf X_3 - i \mathbf X_4 & - \mathbf X_1 + i \mathbf X_2 \\ -2 i \Omega & 0 & - \mathbf X_1 - i \mathbf X_2 & \mathbf X_3 - i \mathbf X_4 \\ -\mathbf X_3 - i\mathbf  X_4 & -\mathbf X_1 - i \mathbf X_2 & 0 & - 2i \Omega \\ -\mathbf X_1 + i \mathbf X_2 &  \mathbf X_3 - i \mathbf X_4 & 2 i \Omega & 0
    \end{pmatrix} \\ & \qquad \qquad = \mathrm{det} \begin{pmatrix}
        2 i \Omega \mathrm{1}_{2(N^2-1)} & H \\ H^\dagger & 2i \Omega  \mathbb{1}_{2 (N^2-1)}
    \end{pmatrix} = \mathrm{det} (H^\dagger H + 4 \Omega^2 \mathbb 1_{2 (N^2-1)}) \,,
}
}
where
\begin{equation}
    H(X) \equiv \begin{pmatrix}
        \mathbf X_1 - i \mathbf X_2 & -\mathbf X_3 - i \mathbf X_4 \\ -\mathbf X_3 + i \mathbf X_4 & - \mathbf X_1 - i \mathbf X_2
    \end{pmatrix} \,.
\end{equation}
Note that $H^\dagger H$ is positive semi-definite so it has eigenvalues $\lambda_i \geq 0$. The determinant is thus of the form
\begin{equation}
    \mathrm{det} (H^\dagger H + 4 \Omega^2 \mathbb 1_{2 (N^2-1)}) = \prod_i (\lambda_i + 4 \Omega^2) \,.
\end{equation}
This is always strictly positive for $\Omega \neq 0$. We thus have
\begin{equation}
    (\mathrm{Pf} \mathcal{M}_{4,\mathrm{I}}(X))^2 = \mathrm{det} \mathcal{M}_{4,\mathrm{I}}(X) > 0
\end{equation}
for all $X$ and $\Omega \neq 0$. Since $\mathrm{Pf} \mathcal{M}_{4,\mathrm{I}}(X)$ is a smooth function of $X$ and $\Omega$, and its square is strictly positive, it is real and does not change sign (for all $\Omega^2 > 0$ and all $X$). Also note that at $\Omega = 0$ it reduces to the massless case which we already proved to be positive semi-definite.
%Thus, the Pfaffian never changes sign when $\Omega \neq 0$. We also know that for $\Omega=0$ it reduces to the Pfaffian of the pure SYM matrix model which we know is positive semi-definite. 
Thus it is positive semi-definite everywhere and we can write
\begin{equation}
\mathrm{Pf} \mathcal{M}_{4,\mathrm{I}}(X) = \sqrt{\mathrm{det} (H^\dagger H + 4 \Omega^2 \mathbb 1_{2 (N^2-1)})} \,.
\end{equation}

\paragraph{Type II:} For the type II theory, it is again convenient to switch to the Weyl representation yielding (in dimensionless coordinates $\mu_i \to \Omega_i$)
\al{
\spl{
\int \left(\prod_{\alpha=1}^2 \prod_{A=1}^{N^2-1} d \chi_{A,\alpha}^* d \chi_{A,\alpha} \right) & \mathrm{exp}\left(-\chi_A^\dagger \left(i f_{ABC}X_I^C \bar \gamma^I - \frac{2}{3} \Omega_1 \delta_{AB}\right) \chi_B\right) \\ & = \det \underbrace{\left(i \bar \gamma^I \otimes \mathbf X_I - \frac{2}{3} \Omega_1 \mathbb{1}_2 \otimes \mathbb 1_{N^2-1}\right)}_{\equiv A_{\text{II}}(X)} \,,
}
}
with 
\begin{equation}
    A_\text{II}(X) = A(X) - \frac{2}{3} \Omega_1 \mathbb 1_{2 (N^2-1)} \,,
\end{equation}
where $A(X)$ was defined in \eqref{eq: Pfaffian A(X) def}. The matrix $A_\text{II}(X)$ satisfies exactly the same property as $A(X)$ \eqref{eq: Pfaffian conjugation property} provided that $\Omega_1$ is real. Thus, one can repeat exactly the same proof as the one used in the undeformed model to show that the Pfaffian is non-negative as long as $\Omega_1 \in \mathbb R$. 

\section{Massless limit calculations}
\label{app: massless limit}
In this appendix, we review the method to approach the $ \Omega =\mu/\sqrt{g_\mathrm{YM}} \to 0$ limit at fixed $N$ that was used in polarized IKKT \cite{Komatsu:2024ydh} and similarly in the BFSS and BMN models \cite{Lin:2014wka,Komatsu:2024vnb}. For now, let us consider the $U(N)$ theory, where we do not impose that the Hermitian matrices be traceless. The case of $SU(N)$ will be discussed afterwards.
In dimensionless units ($\mu \to \Omega, g_\mathrm{YM} \to 1$), the actions we considered in this paper take the form\footnote{The generalization to $D=6$ is straightforward. Note that when the model has more than one parameter, one should think of $\Omega$ as being the scale common to all $\Omega_i$. Namely, one should write $\Omega_i = \epsilon_i \Omega$ with $\epsilon_i \sim \mathcal{O}(1)$ and only then take $\Omega \to 0$.}
\al{
\spl{
S = \mathrm{Tr} \Biggl[ -\frac{1}{4} & [X_I,X_J]^2 - \frac{i}{2} \bar{\psi}\Gamma^{I} [X_I,\psi]  
\\ &  + \Omega S_{I J K} X^I X^J X^K + \Omega^2 S_{I J} X^{I} X^J + \Omega \bar \psi M \psi \Biggr] \,.
}
}
We start by rescaling $X = \frac{1}{\Omega}  X'_I$ and $\psi = \frac{1}{\sqrt{\Omega}} \psi'$. This gives
\al{
\spl{
S' = \mathrm{Tr} \Biggl[ -\frac{1}{4 \Omega^4} & [X'_I,X'_J]^2 - \frac{i}{2 \Omega^2} \bar{\psi}' \Gamma^{I}[X_I',\psi']  
\\ & + \frac{1}{\Omega^2} S_{I J K} X'^I X'^J X'^K + S_{I J} X'^{I} X'^J + \bar \psi' M \psi' \Biggr] \,.
}
}
Then, we separate the matrices as
\begin{equation}
    X'^I = r^I + q^I \,, \quad \quad \psi'_\alpha = \theta_\alpha + \Theta_\alpha \,,
\end{equation}
where $r^I$, $\theta_\alpha$ are diagonal and $q^I$, $\Theta_\alpha$ are off-diagonal. We will use the notation
\begin{equation}
    r_{ab}^I\equiv r_a^I-r_b^I \,, \quad \hat{r}_{a b}^I \equiv r_{a b}^I/|r_{a b}| \,, \quad |r_{a b}| \equiv \sqrt{\sum_I r_{a b}^I r_{a b}^I} \,,
\end{equation}
where $r_a^I$ ($a=1,\ldots,N$) are the (diagonal) components of $r^I$. We then use a $SO(D)$ preserving gauge fixing $\sum_I \hat r_{a b}^I q_{a b}^I\overset{!}{=} 0$ for all pairs $a\neq b$. Namely, at this point we have
\begin{equation}
    Z_{U(N)} = c_N \left(\frac{1}{\Omega^2}\right)^{N^2} \int [dr^I][d \theta_\alpha] [dq^I] [d \Theta_\alpha] \delta(\hat r_{a b} \cdot q_{a b}) (\Delta(r) + \mathcal{O}(\Omega)) e^{- S'} \,,
\end{equation}
where $\Delta(r) = \prod_{a \neq b} |r_{a b}|$ is the Vandermonde determinant, which is the leading term in the Jacobian of the gauge fixing, and $c_N = \frac{1}{N!} \frac{\mathrm{Vol} \ U(N)}{\mathrm{Vol} \  U(1)^N} = (2 \pi)^{(N^2-N)/2} / \prod_{k=1}^N (k!)$ is a volume factor. Details on this gauge fixing method are given in \cite{Komatsu:2024vnb,Lin:2014wka}. 
Then, we rescale 
\begin{equation}
    q^I = \Omega^2 y^I \,, \quad \quad \Theta_\alpha = \Omega \chi_\alpha \,,
\end{equation}
and expand the action $S'$ in powers of $\Omega$. This reduces the partition function to
\begin{equation}
    Z_{U(N)} = c_N \left( \frac{1}{\Omega^2} \right)^N \int [dr^I][d \theta_\alpha] [dy^I] [d \chi_\alpha] \delta(\hat r_{a b} \cdot y_{a b}) \Delta(r) e^{-S_0 - S_{y, \chi}} + \text{subleading} \,,
\end{equation} 
where
\begin{equation}
    S_{y, \chi} \equiv \frac{1}{2} |r_{a b}|^2 y_{a b}^{I} y_{b a}^{I} - \frac{i}{2} r_{a b}^I \bar \chi_{b a} \Gamma^I \chi_{a b} \,,
\end{equation}
\begin{equation}
    S_{0} \equiv S_{I J} r^I_a r^J_a +  \bar \theta_a M \theta_a \,.
\end{equation}
The integrals over $\chi_\alpha$ and $y^I$ reduce to Gaussian integrals. In all dimensions $D=3,4,6,10$ we obtain
\begin{equation}
    \int [dy^I] \delta(\hat r_{a b} \cdot y_{a b}) e^{- \frac{1}{2} |r_{a b}|^2 y_{a b}^I y_{b a}^I} = (2 \pi)^{(N^2-N)(D-1)/2} \prod_{a \neq b} |r_{a b}|^{- (D-1)} \,,
\end{equation}
\begin{equation}
    \int [d \chi_\alpha] e^{\frac{i}{2} r_{a b}^I \bar \chi_{b a} \Gamma^I \chi_{a b}} = \prod_{a \neq b} |r_{a b}|^{\mathcal{N}/2} \,.
\end{equation}
Combined with $\Delta(r)$, the $|r_{a b}|$ factors cancel because $1 - (D-1)+ \mathcal{N}/2 = 0$. Putting all factors together and going back to the original coordinates $r \to \Omega r, \theta \to \sqrt{\Omega} \theta$, we obtain
\begin{equation}
    Z_{U(N)} \xrightarrow{\Omega \to 0} \frac{(2 \pi)^{D(N^2-N)/2}}{\prod_{k=1}^N(k!)} \int\prod_{a=1}^N \left( \prod_{a=1}^D dr_I^a \prod_{\alpha=1}^{\mathcal{N}}d\theta_\alpha^a \right) \mathrm{exp}(-S_\mathrm{diag}) \,,
\end{equation}
where $S_\mathrm{diag}$ is nothing but the original action restricted to diagonal matrices,
\begin{equation}
     S_\mathrm{diag} =\Omega^2 S_{I J} r_I^a r_J^a - \Omega \bar \theta^a_\alpha M_{\alpha \beta}  \theta^a_\beta \,.
\end{equation}
Of course this analysis applies to correlators, yielding, for a gauge invariant function $f$,
\begin{equation}
    \langle f(X_I, \psi_\alpha) \rangle \xrightarrow{\Omega \to 0} \langle f(r_I, \theta_\alpha) \rangle_\mathrm{diag}  \,,\label{eq: observables massless limit}
\end{equation}
where $r_I, \theta_\alpha$ are the diagonal parts of $X_I$, $\psi_\alpha$ and the expectation value is taken with respect to $S_\mathrm{diag}$. Note that all these observables diverge as $\Omega \to 0$. 

Let us finally discuss $SU(N)$ observables. In our normalization, $Z_{SU(N)}$ is simply given by $Z_{U(N)}/Z_{U(1)}$. For correlators, we have the relation
\al{
\spl{
\langle \mathrm{Tr} X^{2 n} \rangle_{U(N)}  &=  \sum_{k=0}^{n} {2 n \choose 2k} \langle \mathrm{Tr} X^{2 k} \rangle_{SU(N)} \left\langle \left(\frac{X}{\sqrt{N}}\right)^{2 n -2k} \right\rangle_{U(1)} \,.
}
}
In particular, if the mass term of some matrix $X$ in the mass-deformed model is $S \supset c \Omega^2 \mathrm{Tr} X^2$, then in the massless limit
\begin{equation}
    \langle \mathrm{Tr} X^{2p} \rangle_{U(N)} \xrightarrow{\Omega \to 0} N\left(\frac{1}{c \Omega^2}\right)^p \frac{\Gamma(p+\frac{1}{2})}{\sqrt{\pi}} \,,
\end{equation}
and thanks to the above relation, we obtain
\begin{equation}
    \langle \mathrm{Tr} X^{2p} \rangle_{SU(N)} \xrightarrow{\Omega \to 0} N\left(\frac{N-1}{c \Omega^2N}\right)^p \frac{\Gamma(p+\frac{1}{2})}{\sqrt{\pi}} \,.
\end{equation}

\section{Gamma matrix conventions}
\label{app: gamma matrix conventions and identities}
In this appendix, we give details on the gamma matrix conventions we have been using in each dimension. We remind the reader that the spinors we consider obey reality conditions inherited from Lorentzian signature. Thus, we write our convention in Lorentzian signature, keeping in mind that the Wick rotation to Euclidean signature is performed by writing $\Gamma^D = i \Gamma^0$, keeping everything else ($\psi,\mathcal{C},B$) unchanged.

\subsection{Ten-dimensional gamma matrices}
In 9+1 dimensions, one can impose a Majorana-Weyl condition on spinors which is consistent with Lorentz transformations. It is convenient to choose a parametrization of the 32 $\times$ 32 gamma matrices that makes it manifest. This is done by first defining 16 $\times$ 16 ``Weyl matrices'' $\gamma^{\mu}$, $\bar \gamma^{\mu}$ as
\begin{equation}
    \gamma^{0} = - \bar \gamma^0 = - \mathbb{1}_{16} \,, \qquad \gamma^I = \bar \gamma^I \,, \quad(I = 1,\ldots,9)
\end{equation}
where $\gamma^{I =1 , \ldots, 9}$ are $16 \times 16$ real and symmetric matrices satisfying $\{ \gamma^{I}, \gamma^J\} = 2 \delta^{I J}$. For an explicit construction, refer to \cite[Eq.3.111]{Freedman:2012zz} or \cite[App.A]{Komatsu:2024ydh}. The $SO(9,1)$ gamma matrices and the chiral matrix $\Gamma_*$ are then given by
\begin{equation}
\Gamma^\mu = \begin{pmatrix}
    0 & \gamma^\mu \\ \bar \gamma^\mu & 0
\end{pmatrix} \,,
\qquad
    \Gamma_* = -\Gamma^0 \Gamma^1\ldots \Gamma^9 = \begin{pmatrix}
    \mathbb{1}_{16} & 0\\0 & - \mathbb{1}_{16} \,
\end{pmatrix} \,.
\end{equation}
There exist two matrices $B$ and $\mathcal{C}$ defined by the conditions
\begin{equation}
    B \Gamma^\mu B^{-1} = (\Gamma^\mu)^* \,, \qquad \mathcal{C} \Gamma^\mu \mathcal{C}^{-1} =- (\Gamma^\mu)^\top  \,.\label{eq: B and C conditions in 10d}
\end{equation}
These matrices are useful for the following properties. If $\xi$ transforms as a spinor, then $B^{-1} \xi^*$ also transforms like a spinor, while $\xi^\top \mathcal{C}$ transforms in the inverse way to $\xi$, so that $\xi_1^\top \mathcal{C} \Gamma^{\mu_1 \ldots \mu_r} \xi_2$ transforms like a $r$-form of $SO(9,1)$. Since $\Gamma_\mu$ are real, $B = \mathbb{1}$. Correspondingly, if $\psi$ transforms as a spinor under Lorentz transformations, $\psi^*$ also transforms as a spinor. It is then consistent to impose the Majorana condition $\psi^* = \psi$ as well as the Weyl condition $\Gamma_* \psi = \psi$. We thus have Majorana-Weyl spinors in 9+1 dimensions, and we choose them to have only their first 16 components non-vanishing. Regarding the matrix $\mathcal{C}$, one can define, up to a phase
\begin{equation}
    \mathcal{C} = i \Gamma^0 = \begin{pmatrix}
        0 & - i \mathbb{1}_{16} \\ i \mathbb{1}_{16}& 0
    \end{pmatrix}  \,.
\end{equation}
The phase is chosen such that the fermionic action $\mathrm{Tr}\{- i\bar \psi \Gamma^\mu [X_\mu, \psi] \}$ would be real in Lorentzian signature, where by reality for anti-commuting fermions $\xi$ and $\eta$ we use the convention $(\xi^\top \eta)^* \equiv \eta^\dagger \xi^* \overset{!}{=} \xi^\top \eta$. Let us now discuss Wick rotation in a bit more detail. We define $\Gamma^{10}$ by Wick rotation such that it satisfies the $SO(10)$ Clifford algebra,
\begin{equation}
    \Gamma^{10} = i \Gamma^0 = \begin{pmatrix}
        0 & - i \mathbb{1}_{16} \\ i \mathbb{1}_{16} & 0
    \end{pmatrix} \, , \qquad \{ \Gamma^I, \Gamma^J \} = 2 \delta^{I J} \,, \qquad (I,J=1,\ldots,10)
\end{equation}
while keeping $B = \mathbb{1}_{32}$ and $\mathcal{C} = i \Gamma^0 = \Gamma^{10}$ unchanged. Notice that this matrix $B = 1$ does \textit{not} satisfy \eqref{eq: B and C conditions in 10d} for $SO(10)$ since $\Gamma^{10}$ is imaginary. The matrix $B$ of $SO(10)$ would be an off-diagonal matrix that mixes left-handed and right-handed spinors. This is the reason why Majorana-Weyl fermions do not exist in Euclidean signature. However, all along this paper, we consider the Euclidean theory only as the \textit{Wick-rotated} Lorentzian theory, since our interest for the Euclidean theory is only for integral convergence purposes. Thus, we write Euclidean gamma matrices while keeping our fermions as Lorentzian Majorana-Weyl spinors.

\subsection{Six-dimensional gamma matrices}
\label{app: conventions in d=6}
In 5+1 dimensions, one can proceed similarly in the $9+1$-dimensional case and parameterize the $8 \times 8$ gamma matrices by first defining $4 \times 4$ Weyl matrices $\gamma^\mu, \bar \gamma^\mu$ which are given by
\begin{equation}
    \gamma^0 = - \bar \gamma^0 = - \mathbb{1}_8 \,, \qquad \gamma^I = \bar \gamma^I \,, \qquad (I=1,\ldots,5)
\end{equation}
where $\gamma^I$ are $SO(5)$ matrices satisfying $\{ \gamma^I, \gamma^J \} = 2 \delta^{I J}$. For concreteness we will choose
\al{
\ga{
\gamma^1 = \sigma^3 \otimes \sigma^1 \,, \qquad \gamma^2 = \sigma^3 \otimes \sigma^2 \,, \qquad \gamma^3 = \sigma^3 \otimes \sigma^3 \,, \\
\gamma^4 = - \sigma^2 \otimes \mathbb{1}_2 \,, \qquad \gamma^5 = \sigma^1 \otimes \mathbb{1}_2 \,. \label{eq: 6d weyl matrices}
}
}
We then define the $SO(5,1)$ gamma matrices and the chiral matrix $\Gamma_*$ as
\begin{equation}
\Gamma^\mu = \begin{pmatrix}
    0 & \gamma^\mu \\ \bar \gamma^\mu & 0
\end{pmatrix} \,,
\qquad
    \Gamma_* = \Gamma^0 \Gamma^1\ldots \Gamma^5 = \begin{pmatrix}
    \mathbb{1}_{4} & 0\\0 & - \mathbb{1}_{4}
\end{pmatrix} \,.
\end{equation}
Now the matrices $B$ and $\mathcal{C}$ satisfying \eqref{eq: B and C conditions in 10d} are given by
\begin{equation}
    B = \Gamma^2 \Gamma^4 \,, \qquad \mathcal{C} = i B \Gamma^0 \,.
\end{equation}
At this point, one can impose a Weyl condition $\Gamma_* \psi = \psi$, which would leave 8 degrees of freedom $\psi_{\alpha = 1,\ldots,4}$ and $\psi^*_{\alpha = 1,\ldots,4}$. However, a Majorana condition $\psi = B^{-1} \psi^*$ cannot be imposed since consistency would require $B^* B = \mathbb{1}_8$ which is not the case here ($B^* B = - \mathbb{1}_8$). To get around this, let us have two Weyl fermions $\psi_1, \psi_2$, together with their conjugates $\psi_1^*$, $\psi_2^*$, and define a symplectic Majorana-Weyl condition,
\begin{equation}
    \psi_a = \epsilon^{a b}B^{-1} \psi_b^* \,,\quad (a,b=1,2) \label{eq: symplectic condition}
\end{equation}
where $\epsilon^{i j}$ is the Levi-Civita symbol with $\epsilon^{1 2} = 1$. Now this condition is consistent thanks to $\epsilon^{i j} \epsilon^{j k} = - \delta^{ik}$ which effectively cancels the minus sign that we had in $B^* B =-1$. This condition allows us to write $\psi_1^*$ and $\psi_2^*$ in terms of $\psi_1, \psi_2$ for a total number of 8 degrees of freedom. This representation of the spinors is the so-called symplectic Majorana-Weyl representation \cite{VanProeyen:1999ni, Kugo:1982bn, Park:1998nra}. Note that it is equivalent to the Weyl representation, since one can use the symplectic condition to write $\psi_2, \psi_2^*$ in terms of a single Weyl spinor and its conjugate $\psi_1, \psi_1^*$.

Finally, the Wick rotation is performed by setting 
\begin{equation}
    \Gamma^6 = i \Gamma^0 \qquad \{ \Gamma^I , \Gamma^J \} = 2 \delta^{I J} \,,\qquad (I,J=1,\ldots,6)
\end{equation}
and keeping $\mathcal{C} = i B \Gamma^0$, $B = \Gamma^2 \Gamma^4$.

\subsection{Four-dimensional gamma matrices}
\label{app: D=4 gamma matrices}
In $3+1$ dimensions one can have either a Majorana or a Weyl condition. In this paper, we have chosen the Majorana condition to classify the $D=4$ models on the same footing as $D=3,10$. Note that the Majorana representation is equivalent to the Weyl representation. We will discuss how to relate the two. 

We choose a convention for the gamma matrices which makes the Majorana condition manifest. Namely, we define the four real $4 \times 4$ gamma matrices as well as the chiral matrix $\Gamma_*$ by
\begin{equation}
    \Gamma^{0} = i \sigma^2 \otimes \mathbb{1}_2 \,, \qquad \Gamma^1 = \sigma^3 \otimes \mathbb{1}_2 \,, \qquad \Gamma^2 = \sigma^1 \otimes \sigma^3 \,, \qquad \Gamma^3 = \sigma^1 \otimes \sigma^1  \,,\label{eq: 4d Majorana Gamma matrices}
\end{equation}
\begin{equation}
    \Gamma_* = i \Gamma^0 \Gamma^1 \Gamma^2 \Gamma^3 \,,
\end{equation}
such that
\begin{equation}
    \{ \Gamma^\mu, \Gamma^\nu \} = 2 \eta^{\mu \nu} \,.
\end{equation}
Since these are real matrices we have $B = \mathbb{1}_4$ and $\psi = \psi^*$ can be imposed. The conjugation matrix $\mathcal{C}$ reads
\begin{equation}
    \mathcal{C} = i \Gamma^0 \,.
\end{equation}
The Wick rotation is performed by setting
\begin{equation}
    \Gamma^{4} = i \Gamma^0 \, , \qquad \{ \Gamma^I, \Gamma^J \} = 2 \delta^{I J} \qquad (I,J=1,\ldots,4)
\end{equation}
while keeping $B = 1$ and $\mathcal{C} = i \Gamma^0 = \Gamma^4$.

Let us now discuss how to map the Majorana representation to the Weyl representation. If one has a real spinor $\psi$ in the above representation, it is always possible to rewrite it in terms of a Weyl spinor $\chi = (\chi_1, \chi_2, 0,0)$ and its conjugate $\chi^*= (\chi_1^*, \chi_2^*, 0,0)$ as
\begin{equation}
    \psi = U \chi + U^* \chi^* \,,
\end{equation}
where $U$ is a $4 \times 4$ unitary matrix chosen such that
\begin{equation}
    U^\dagger \Gamma_* U = \begin{pmatrix}
        \mathbb{1}_2 & 0 \\ 0 & - \mathbb{1}_2
    \end{pmatrix} \,, \qquad U^\dagger \Gamma^\mu U = \begin{pmatrix}
        0 & \gamma^\mu \\ \bar \gamma^\mu & 0
    \end{pmatrix} \,,
\end{equation}
where $\gamma^\mu, \bar \gamma^\mu$ are the Weyl matrices
\begin{equation}
    \gamma^0 = - \bar \gamma^0 = - \mathbb{1}_2 \,,\qquad \gamma^i = \bar \gamma^i = \sigma^i \,, \label{eq: gbar in 4d}
\end{equation}
with $\sigma^i$ the Pauli matrices. Explicitly, with the matrices \eqref{eq: 4d Majorana Gamma matrices}, $U$ is given by
\begin{equation}
    U = \frac{e^{i \pi /4}}{2}\begin{pmatrix}
        -1 & - i & -i & -1 \\ 1 & - i & - i & 1 \\ -i & -1 & 1 & i \\ -i & 1 & -1 & i
    \end{pmatrix} \,.\label{eq: U from Maj to Weyl}
\end{equation}

\subsection{Three-dimensional gamma matrices}
In $2+1$ dimensions, since the dimension is odd, there are no Weyl fermions. One can however impose a Majorana condition. To make it manifest, we write the gamma matrices as
\begin{equation}
    \Gamma^0 = \begin{pmatrix}
        0 & -1 \\ 1 & 0
    \end{pmatrix} \,,\qquad \Gamma^1 = \begin{pmatrix}
        0 & 1 \\ 1 & 0
    \end{pmatrix}\,,  \qquad \Gamma^2 = \begin{pmatrix}
        1 & 0 \\ 0 & -1
    \end{pmatrix}  \,.
\end{equation}
Hence, $B = \mathbb{1}_2$ and we can impose $\psi^* = \psi$. The conjugation matrix $\mathcal{C}$ is set to
\begin{equation}
    \mathcal{C} = i \Gamma^0 \,.
\end{equation}
The Wick rotation is performed by setting
\begin{equation}
    \Gamma^3 = i \Gamma^0 = \begin{pmatrix}
        0 & - i \\ i & 0
    \end{pmatrix} \,, \qquad \{ \Gamma^I, \Gamma^J \} = 2 \delta^{I J} \,, \qquad (I,J =1,\ldots,3)
\end{equation}
while keeping $\mathcal{C} = i \Gamma^0 = \Gamma^3$ and $B = \mathbb{1}$ fixed.

\section{Matrix integral conventions}
\label{app: matrix integral conventions}
Here, we give details on the normalization conventions we use for the matrices and the integral measures. These conventions are the same as the one used in \cite{Komatsu:2024ydh}. We expand our Hermitian (traceless) $N \times N$ matrices $X_I,\psi_\alpha$ in terms of the $U(N)$ ($SU(N)$) generators in the defining representation. 
\begin{equation}
    X_I = X_I^A T^A \,, \qquad \psi_\alpha = \psi_\alpha^A T^A \,.
\end{equation}
The generators $T^A$ are defined as a basis of $N \times N$ Hermitian (traceless) matrices satisfying
\begin{equation}
    \mathrm{Tr} T^A T^B = \delta^{AB} \,.
\end{equation}
Given these generators, one defines the structure constants $f_{ABC}$ from $[T^A,T^B] = if^{ABC}T^C$.
The measure is defined as
\begin{equation}
    [d X_I] \equiv \prod_{A=1}^{N^2} dX_I^A \,,\qquad [d \psi_\alpha] \equiv \prod_{A=1}^{N^2} d \psi_\alpha^A \,, \qquad (U(N))
\end{equation}
\begin{equation}
    [d X_I] \equiv \prod_{A=1}^{N^2-1} dX_I^A \,,\qquad [d \psi_\alpha] \equiv \prod_{A=1}^{N^2-1} d \psi_\alpha^A \,.\qquad (SU(N))
\end{equation}
The reason why we give our preference to this convention is that the measure satisfies the nice properties
\begin{equation}
    \int [d X] e^{- \lambda \mathrm{Tr} X^2}  = \begin{cases}
        \left(\frac{\pi}{\lambda}\right)^{N^2/2} & (U(N)) \\ 
        \left(\frac{\pi}{\lambda}\right)^{(N^2-1)/2} & (SU(N)) \\ 
    \end{cases} \,, \label{eq:BosonicGaussian}
\end{equation}
\begin{equation}
    \int [d \eta] [d \psi]  e^{- \lambda \mathrm{Tr} \eta \psi} = \begin{cases}
        \lambda^{N^2} & (U(N)) \\ \lambda^{N^2-1} & (SU(N))
    \end{cases} \,, \label{eq:FermionicToyGaussian}
\end{equation}
as if these were simple $N^2$ or $N^2-1$ bosonic and fermionic Gaussian integrals.

\bibliographystyle{utphys} % We choose the "plain" reference style
\bibliography{refs}

\end{document}